\documentclass[acmtog]{acmart}

\hypersetup{draft}

\usepackage{soul}
\usepackage{mystyle}
\usepackage{wrapfig}

\acmJournal{TOG}
\acmVolume{39}
\acmNumber{3}
\acmArticle{25}
\acmYear{2020}
\acmMonth{4}

\setcopyright{rightsretained}

\begin{document}

\acmJournal{TOG}
\acmYear{2020} \acmVolume{1} \acmNumber{1} \acmArticle{1} \acmMonth{1} \acmPrice{}\acmDOI{10.1145/3374209}

\title{Octahedral Frames for Feature-Aligned Cross-Fields}

\author{Paul Zhang}
\orcid{0000-0003-4136-1315}
\affiliation{%
  \institution{Massachusetts Institute of Technology}
  \streetaddress{77 Massachusetts Avenue}
  \city{Cambridge}
  \state{MA}
  \postcode{02139}
  \country{USA}}
\email{pzpzpzp1@mit.edu}

\author{Josh Vekhter}
\affiliation{%
  \institution{University of Texas at Austin}
  \city{Austin}
  \state{TX}
  \postcode{78712}
  \country{USA}}
\email{josh@cs.utexas.edu}

\author{Edward Chien}
\affiliation{%
  \institution{Massachusetts Institute of Technology}
  \streetaddress{77 Massachusetts Avenue}
  \city{Cambridge}
  \state{MA}
  \postcode{02139}
  \country{USA}}
\email{eddchien@csail.mit.edu}

\author{David Bommes}
\affiliation{%
  \institution{University of Bern}
  \streetaddress{Hochschulstrasse 6}
  \city{Bern}
  \postcode{3012}
  \country{Switzerland}}
\email{david.bommes@inf.unibe.ch}

\author{Etienne Vouga}
\affiliation{%
  \institution{University of Texas at Austin}
  \city{Austin}
  \state{TX}
  \postcode{78712}
  \country{USA}}
\email{evouga@cs.utexas.edu}

\author{Justin Solomon}
\orcid{0000-0002-7701-7586}
\affiliation{%
  \institution{Massachusetts Institute of Technology}
  \streetaddress{77 Massachusetts Avenue}
  \city{Cambridge}
  \state{MA}
  \postcode{02139}
  \country{USA}}
\email{jsolomon@mit.edu}

\begin{abstract}
We present a method for designing smooth cross fields on surfaces that automatically align to sharp features of an underlying geometry.  Our approach introduces a novel class of energies based on a representation of cross fields in the spherical harmonic basis.  We provide theoretical analysis of these energies in the smooth setting, showing that they penalize deviations from surface creases while otherwise promoting intrinsically smooth fields.  We demonstrate the applicability of our method to quad-meshing and include an extensive benchmark comparing our fields to other automatic approaches for generating feature-aligned cross fields on triangle meshes.  
\end{abstract}



\begin{teaserfigure}
  \centering
  \includegraphics[width=\textwidth,keepaspectratio]{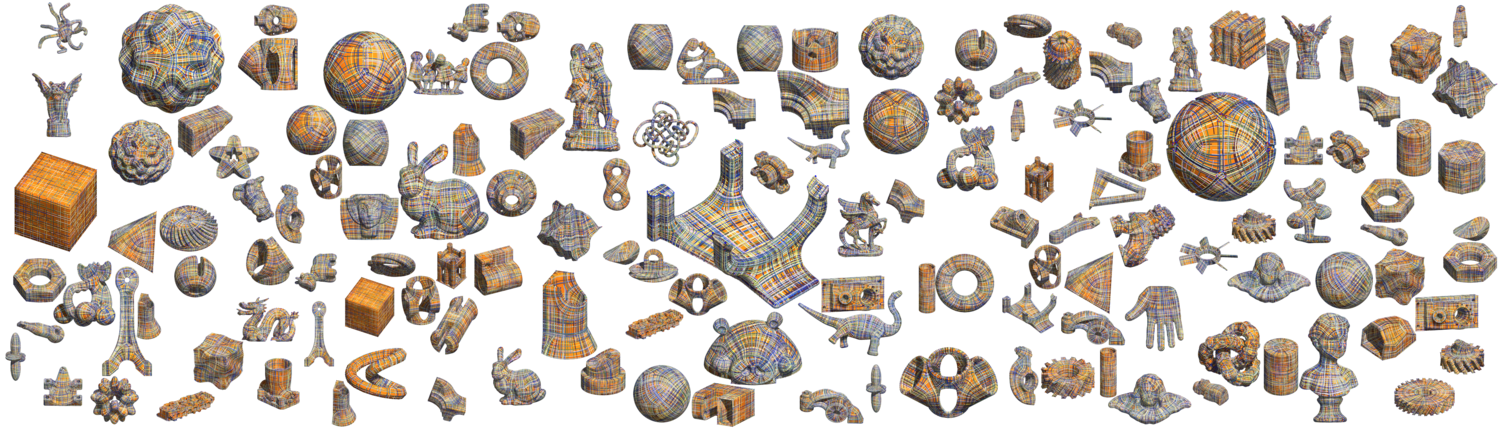}
  \caption{A variety of feature-aligned cross fields computed using our novel cross field formulation.}
  \label{fig:mosaic_teaser}
\end{teaserfigure}

\maketitle

\section{Introduction}

$N$-rotationally symmetric (RoSy) tangential vector fields over surfaces are ubiquitous in computer graphics. 2-RoSy fields can be used to generate stripe patterns  
due to their ambivalence to rotation by $\pi$ about the normal. 4-RoSy fields (cross fields) are heavily used in both surface parameterization and quadrilateral (quad) meshing, thanks to their symmetry with respect to rotations by $\frac{\pi}{2}$ about the surface normal. 

Depending on the application, $n$-RoSy field design algorithms must trade off between several desirable properties of the field. In almost all cases, $n$-RoSy fields are expected to be as smooth as possible. For surfaces with boundary, constraints on how the field aligns to the boundary are common, and for artistic applications,  users may wish to  prescribe a sparse set of streamlines that the field must follow. For meshing  applications,  alignment of $n$-RoSy fields to salient geometric features is also desirable as a means to identify or preserve mesh detail. Our focus will be on improving this latter aspect for the important case of $4$-RoSy fields.

There are two broad strategies for achieving feature alignment. The first is to optimize only for smoothness, under the assumption that a well-chosen functional for measuring cross-field smoothness will automatically penalize fields that fail to align to geometric features.  The most commonly-used smoothness functionals (including the Dirichlet energy and its variants) are \emph{intrinsic}, and recover solutions that are unique only up to rotation \citep{knoppel_globally_2013}. These are ambivalent to isometric deformations of the surface and ignore \emph{extrinsic} features such as creased folds.

\begin{figure}[t]
    \vspace{-15pt}
    \centering
    \includegraphics[width=.45\columnwidth]{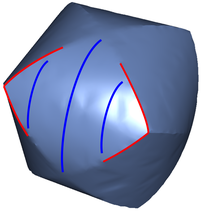}\;\;
    \includegraphics[width=.45\columnwidth]{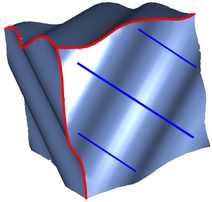}
    \vspace{-12pt}
	\caption{Two surfaces (the \textbf{three-cylinder-intersection} and \textbf{wavy-box}) whose maximal curvature directions (blue lines)  contradict its feature curves (red lines).}
	\label{fig:counterExamples}
\end{figure}

An alternative strategy is to include energy terms that explicitly enforce alignment to an input guiding field of principal curvature directions during cross-field design~\cite{knoppel_globally_2013,brandt_modeling_2018}. Drawbacks include the difficulty of robustly computing principal curvature directions on noisy meshes, the fact that forcing alignment to a guiding field based on local geometry may exclude cross-field designs that are globally more optimal, and more critically the fact that principal curvature directions are often different from features e.g. Figure \ref{fig:counterExamples}. 

Our main observation is that neither of the above strategies adequately identify those features most important to generating high-quality quad meshes.
Often the surface being modelled is constructed from smooth patches that are joined along \emph{sharp extrinsic feature curves} where the normal direction is discontinuous or changes rapidly. On the one hand, such features are invisible to intrinsic smoothness functionals; on the other, the orientation of the feature curves often contradict that of nearby curvature lines.

Consider the surfaces shown in Figure~\ref{fig:counterExamples}: Neither existing strategy will promote alignment to the features curves shown in red. Both of these shapes are developable away from a sparse set of cone singularities at the corners; the gaussian curvature is nearly zero at creased edges and curved facets, and so purely intrinsic approaches have no hope of aligning to the creases.  Augmenting with a guiding field based on extrinsic curvature is counter-productive, as the curvature lines (blue) are not compatible with the surface's more important crease features curves (red).

We approach feature alignment in a new way, which detects and aligns cross fields to sharp features in a stable fashion. Our method is based on an extrinsic representation of cross fields using spherical harmonic (SH) basis functions. SH functions have been used successfully in volumetric octahedral field problems for hexahedral meshing \cite{solomon_boundary_2017,huang_boundary_2011,ray_practical_2016}, and we argue that this representation is well-suited not only for computing octahedral fields in volumes but also for field computation on surfaces. In particular, we apply an SH representation of octahedral frames, or frames of three orthogonal directions in $\R^3$, proposed by \citet{huang_boundary_2011}; when one of its directions is constrained to the surface normal, it exhibits the same symmetry as a two-dimensional cross. We use this fact to devise a class of cross field energies that promote intrinsic smoothness in smooth regions of the surface. Over sharp creases, however, our energy aligns the field to the crease direction, achieving automatic feature alignment without the need for explicit computation of extrinsic curvature directions or feature curves. 

\paragraph{Contributions}
In this work, we
\begin{itemize}
	\item introduce spherical harmonic functions for the computation of surface cross fields;
	\item propose a family of field smoothness energies whose optima are feature-aligned cross fields;
	\item provide a theoretical analysis of the behavior of a few important members of this family; and 
	\item introduce cross fields with soft normal alignment for increased versatility/robustness.
\end{itemize}
Our approach is able to extract feature-aligned fields with comparable levels of efficiency to those of purely intrinsic algorithms. We test our algorithm extensively on over 200 different meshes with results in both \S \ref{sec:results} and in supplementary materials.  We leverage our algorithm to produce feature-aligned cross fields and demonstrate their usefulness for quad meshing.

\section{Related Work}

The generation of tangential $n$-RoSy fields over surfaces has many applications in computer graphics ranging from surface BRDF modification~\cite{brandt_modeling_2018} to meshing~
\cite{jakob_instant_2015,bommes_mixed-integer_2009}
to~texture synthesis \cite{knoppel_stripe_2015} and sketch-based modeling~\cite{iarussi_bendfields:_2015,bessmeltsev_vectorization_2018}.
Surveys of $n$-RoSy field design methods are provided in \cite{vaxman_directional_2016} and \cite{doGoes:2015:VFP:2818143.2818167}.

\vspace{-2pt}
\subsection{Cross Field Design}
Cross fields ($n=4$) have been especially well-studied since their $\frac{\pi}{2}$-symmetry allows them to behave like local coordinate systems, resulting in intuitive seamless surface parameterizations.

Methods to compute intrinsically-smooth cross fields with alignment and singularity constraints were studied by \citet{Ray:2008:NDF:1356682.1356683}, \citet{crane_trivial_2010} and \citet{knoppel_globally_2013}. More similar to our work, \citet{jakob_instant_2015} instead formulate an \emph{extrinsic} smoothness functional on cross fields, in an attempt to automatically align to surface features. Their method penalizes an extrinsic distance between neighboring crosses that does not use a shared tangent space or connection.
The resulting energy is non-convex but is minimized to local optimality, often resulting in more singularities than necessary.  \citet{huang_extrinsically_2016} analyze this extrinsic energy, and find that it can be decomposed into an energy expressed in terms of intrinsic twisting and alignment to extrinsic curvature directions.  
We perform a similar analysis of the energy we introduce in the supplemental documents.

When using cross fields for quad mesh parameterization or processing, methods~\cite{bommes_mixed-integer_2009, campen_scale-invariant_2016,knoppel_globally_2013,brandt_modeling_2018} often promote feature alignment by including a loss term penalizing disagreement with curvature directions. However, as we argue in the introduction and illustrate in Figure~\ref{fig:counterExamples}, alignment to curvature directions is often less important than alignment to sharp creases. Other parameterization methods such as \cite{bommes_mixed-integer_2009,bommes:hal-integer-grid-maps,campen_quantized_2015} allow feature alignment but just assume that such feature curves are provided as input.

\vspace{-5pt}
\subsection{Octahedral Fields and Volumetric Representations}

The three-dimensional generalization of a cross field is an \emph{octahedral field}. Octahedral fields are often used in volumetric problems like hexahedral meshing \cite{nieser_cubecover-_2011}. A single octahedral frame consists of three mutually-orthogonal vectors and their negations. \citet{huang_boundary_2011} introduced a particularly convenient representation of an octahedral frame as a rotation of the spherical function $g(x,y,z):(x,y,z)\in S^2\mapsto x^4+y^4+z^4$, encoded by coefficients in the spherical harmonic (SH) basis. \citet{ray_practical_2016} used this representation to generate volumetric normal-aligned octahedral fields, and \citet{solomon_boundary_2017} combined the SH representation with the boundary element method to remove the need for a volumetric mesh.  Both of these methods use normal-alignment constraints at the surface to enforce alignment of the octahedral frame with the volume's boundary.

While there is no 
canonical three-dimensional generalization of arbitrary $n$-RoSy fields, the spherical harmonic representation allows for frames that mimic the symmetries of all platonic solids~\cite{shen_harmonic_2016}, including octahedral fields
~\cite{solomon_boundary_2017,liu_singularity-constrained_2018,corman_symmetric_2019}.
Algebraic characterization of the orbit of $g(x,y,z)$ under the space of rotations as a subset of all possible SH coefficients was presented by \citet{anonymous_algebraic_2019} and \citet{chemin_representing_2018}. 

\vspace{-5pt}
\section{Preliminaries}
\label{sec:preliminaries}

 Since our formulation relies heavily on both the spherical harmonic representation of octahedral frames and vectorial total variation, we present a preliminary introduction to these topics.
\vspace{-5pt}
\subsection{Spherical Harmonic (SH) Octahedral Frames}\label{sec:sphframes}

\begin{wrapfigure}[8]{r}{0.40\columnwidth}
\vspace{-12pt}
  \hspace*{-20pt}
    \includegraphics[draft=false,width=0.45\columnwidth]{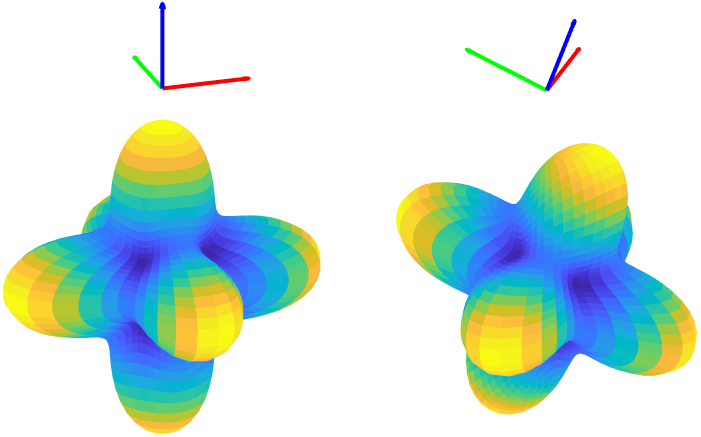}
    \put(-55,63){$e^{[v]}$}
    \put(-59,58){$\longrightarrow$}
    \put(-59,10){$\longrightarrow$}
    \put(-57,2){$e^{v\cdot L}$}
    \put(-105,-10){rotation of octahedral frame}
  \vspace{-4pt}
  \captionsetup{labelformat=empty}
\end{wrapfigure}

As introduced by \citet{huang_boundary_2011}, the canonical axis-aligned octahedral frame can be represented by spherical harmonics as a function $g_0:\mathbb{S}^2\to \R$ written as $g_0=\sqrt{\frac{5}{12}}Y_{44} + \sqrt{\frac{7}{12}}Y_{40},$ where $Y_{lm}$ denotes the basis for real spherical harmonics. 
The function $g_0$ can be understood as the scaled projection of $x^4+y^4+z^4$ onto the fourth band ($l=4$) of spherical harmonics. Written differently, we can encode $g_0$ as a vector of coefficients in the full basis of fourth-band spherical harmonics $Y_{4(-4)},\ldots,Y_{44}$: 
$$f_0=\left[0,0,0,0,\sqrt{\frac{7}{12}},0,0,0,\sqrt{\frac{5}{12}}\right]^T.$$ 

The space of octahedral frames can be described as all rotations of the canonical octahedral frame, that is, the orbit of $f_0$ under the group of 3D rotations $\mathrm{SO}(3)$ 
\cite[Definition 3.5]{anonymous_algebraic_2019}. 
We write this via exponentiation of the Lie algebra elements: the set of octahedral frames is 
$$\mathcal{V} = \left\{f\ \Big|\ \textrm{there exists }{v\in \R^3} \textrm{ with } f = e^{v \cdot L} f_0\right\},$$
where $v\cdot L = v_x L_x + v_y L_y + v_z L_z$ and $L_x,L_y,L_z$ are the angular momentum operators expressed in the basis of band-four spherical harmonics. In this basis, $L_x,L_y,L_z$ are each $9\times 9$ matrices. The angular momentum operators are explicitly written in supplementary materials, \S4. In this description, $v$ can be interpreted as an axis-angle representation of rotation, with corresponding rotation matrix $e^{[v]}$. 
$[v]$ denotes the skew-symmetric matrix that acts as $[v]u=v\times u$. Accordingly, $e^{v\cdot L}g_0$ encodes the octahedral frame whose directions are $\hat{x},\hat{y},\hat{z}$ rotated by $e^{[v]}$, where 
$\hat{ }$
denotes normalization (see inset above).

\begin{wrapfigure}[8]{r}{0.2\textwidth}
\vspace{-10pt}
\hbox{
\hspace{-15pt}
  \centering
    \includegraphics[width=0.3\textwidth]{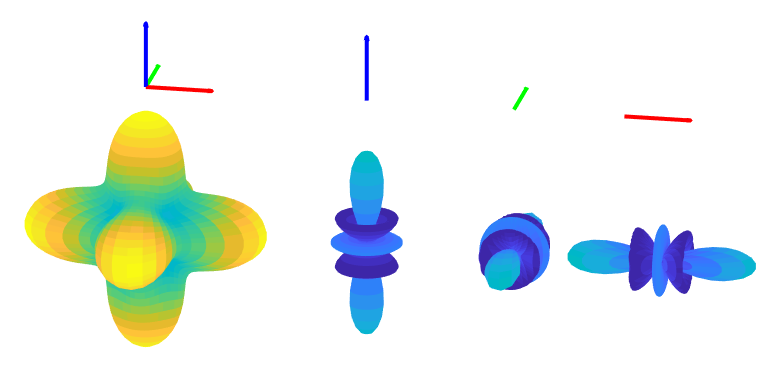}
    \put (-45,35) {$+$}
    \put (-68,35) {$+$}
    \put (-98,35) {$=$}
    }
    \vspace{-10pt}
  \captionsetup{labelformat=empty}
  \caption{SH frame as sum of three lobes}
\end{wrapfigure}
Using such SH rotations, we can present an alternative interpretation of the octahedral frame $f_0$ as the sum of three orthogonal SH lobe-shaped functions. The $z$-aligned lobe is $l=[0,0,0,0,\sqrt{\frac{7}{12}},0,0,0,0]$ and is depicted in the inset. Lobes can be rotated in the same way that frames can i.e. by applying $e^{v\cdot L}$. 
The canonical octahedral frame can therefore be equivalently expressed as $f_0 = l + e^{\frac{\pi}{2}L_x}l + e^{\frac{\pi}{2}L_y}l$.

The space of octahedral frames that are aligned to a unit vector $\hat{n}$ can be described by the set
$$
\left\{ e^{v_n \cdot L} e^{\theta L_z} f_0\ \Big|\ \theta \in S^1\right\},$$ 
where $v_n$ is any axis-angle rotation taking $\hat{z}$ to $\hat{n}$, (e.g., the vector parallel to $\hat{z} \times \hat{n}$ and has magnitude equal to the angle from $\hat{z}$ to $\hat{n}$) and $\theta$ encodes an additional twist of the frame about $\hat{n}$. The first rotation about $\hat{z}$ can be written in explicit form~\cite{huang_boundary_2011} as
$$e^{\theta L_z} f_0=\left[\sqrt{\frac{5}{12}} \; \cos 4\theta ,0,0,0, \sqrt{\frac{7}{12}},0,0,0,\sqrt{\frac{5}{12}} \; \sin 4\theta \right]^T.$$
The above allows us to formulate the set of all octahedral frames $g$ aligned to a given direction $\hat{n}$ in terms of two constraints:
\begin{equation}\label{eq:normalAlignment}
\|f\|_2=1,\;\;W_n f = u_0=\left[0,0,0,\sqrt{\frac{7}{12}},0,0,0\right]^T,    
\end{equation}
where $W_n$ is the second through eighth rows of $e^{-v_n \cdot L}$. The linear constraint rotates the frame from normal alignment to $\hat{z}$ alignment, and the norm constraint ensures that the first and last components are of the appropriate form.

Lastly, we will make use of the projection operator $\pi_{\mathcal{V}}: \R^9 \rightarrow \mathcal{V}$ onto the space of octahedral frames $\mathcal{V}$, as defined in \cite[\S\ref{sec:nontriviality}]{anonymous_algebraic_2019}.

\subsection{Vectorial Total Variation}
We will later make use of a total variation energy (amongst others) to analyze the behavior of our cross fields on creased surfaces. Here, we
introduce total variation and vectorial total variation definitions in $\R^n$ and provide intuition about their use. The extension to functions on a Riemannian manifold is straightforward, using the standard intrinsic gradient and divergence operators.

The total variation of a differentiable scalar function $h:\Omega\rightarrow \R$ is $TV[h]=\int_{\Omega}\|\nabla h\|_2~dA$ where $\Omega\subset \R^n$ \cite{ambrosio_functions_2000}. For non-differentiable $h$, the relevant definition is: 
\begin{equation*}
    TV[h] = \sup_{\phi\in C^1_c, \forall_x \|\phi(x)\|_2\leq1}\left(\int_{\Omega} h \nabla \cdot \phi~dA\right),
\end{equation*}
where $C^1_c$ denotes differentiable, compactly-supported vector fields. For smooth $h$, equivalence to $\int_{\Omega}\|\nabla h\|$ follows from integration by parts and Stokes's theorem. In this case, the maximizing $\phi$ is $\nicefrac{-\nabla h}{\|\nabla h\|}$. If $h$ is the indicator function of a suitably regular (e.g., non-fractal) subset $A\subset \Omega$, then $TV[h]$ is the perimeter of $A$. 

When $h:\Omega\rightarrow \R^m$ is vector-valued rather than scalar-valued, there are many different definitions for the vectorial total variation $VTV[h]$ \cite{sapiro_vector-valued_1996}. We use one proposed by \citet{di_zenzo_note_1986}, which for differentiable $h$ is given by $VTV[h]=\int_{\Omega}\|\nabla h\|_F~dA$, where $\|\cdot\|_F$ is the Frobenius norm.  More generally, we can take
\begin{equation}\label{eq:vtv}
    VTV[h] = \sup_{\phi\in C^1_c, \forall_x \|\phi(x)\|_F\leq1}\left(\sum_{i=1}^m \int_{\Omega} h_i \nabla \cdot \phi_i \right),
\end{equation}
where $h=(h_1,h_2,\ldots,h_m)$, and $\phi = (\phi_1, \phi_2, \ldots \phi_m)$ is a differentiable, compactly-supported $m$-tuple of vector fields. This definition is \emph{not} equivalent to a sum of $m$ independent scalar total variations: The constraint on $\phi$ introduces nontrivial coupling between the dimensions. This definition of total vectorial variation is considered in the case where $\Omega$ is a surface in $\R^3$ by \citet{bresson_fast_2008}, but without specific analysis for discontinuous $h$. 

\section{Spherical Harmonic Octahedral Frames as Cross Fields}
We use normal-aligned octahedral fields to encode tangent cross fields on surfaces, with the goal of computing a smooth cross field on a surface aligned to sharp features. The SH representation will enable us to capture features even when they are purely extrinsic. To this end, our next task is to define a means of measuring smoothness by examining the gradient of a SH field along the surface.

\subsection{Derivatives of SH Octahedral Frames}

To calculate $\|\nabla f\|^2$, we first express it in an appropriate local coordinate system that simplifies the formulas in coordinates and better reveals the structure.  Following the
 notation in~\S\ref{sec:sphframes}, an octahedral field $f(r):\Omega \rightarrow \mathcal{V} \subset \R^9$ can be parameterized relative to a point $r^*$ by $v(r):\Omega \rightarrow \R^3$, where $v(r)$ is the axis-angle rotation from $f(r^*)$ to $f(r)$. This implies that $v(r^*)=[0,0,0]$. Without loss of generality, we rotate the surface so that the normal of $\Omega$ at $r^*$ is $\hat{z}$. We can then compute the gradient $\nabla f$ at the point $r^*$ from the formula $f(r)=e^{v(r)\cdot L}f(r^*)$:

\begin{equation}
\hspace{-10pt}
\label{eq:gradSPH}
\begin{aligned}
    \nabla f(r)|_{r^*} = \!\!
    \begin{bmatrix}
        | & | & |\\
        L_x f(r^*) & \!\!L_y f(r^*) & \!\!L_z f(r^*)\\
        | & | & |
    \end{bmatrix}
    \begin{bmatrix}
        | & | & |\\
        \nabla_x v & \!\!\nabla_y v & \!\!\nabla_z v\\
        | & | & |
    \end{bmatrix}_{r^*}.
\end{aligned}
\end{equation}

As the field $f(r)$ encodes an extrinsically embedded frame at each point, we take the gradient $\nabla$ to be the component-wise derivative of the field's nine scalar functions rather than a covariant or Lie derivative along the surface in order to capture the extrinsic geometry of the surface.
We use \citep[\S 1.2.5]{rossmann_lie_2002} and the fact that $v(r^*)=[0,0,0]$ to derive Equation~\eqref{eq:gradSPH}. 

By combining facts about the SH representation and standard results in differential geometry we show that the squared norm $\|\nabla f(r)\|^2$ at $r^*$ can then be expressed in the following more intuitive way:
\begin{proposition}
	\label{prop:txty}
	Let $f(r):\Omega\rightarrow\mathcal{V}\subset\R^9$ be a normal-aligned octahedral field over a smooth surface $\Omega$. 
	Then at every point $r^* \in \Omega$, $\|\nabla f\|^2_F = k_1^2 + k_2^2 + w$, where $k_1$ and $k_2$ are the principal curvatures and $w$ measures the intrinsic tangential twist of the octahedral field. 
	Using mean and Gauss curvatures $H$ and $K$, we can write $\|\nabla f\|^2_F = 2H^2-K+w$.
\end{proposition}
We leave the full proof of this formula to supplementary materials. Proposition \ref{prop:txty} gives a more intuitive form for Equation~\eqref{eq:gradSPH} and relates the spherical harmonic representation of an octahedral frame to properties of the frame it represents. Most notably, the Dirichlet energy of the SH representation can be effectively decoupled into extrinsic dependence of $\|\nabla f\|^2_F$ on the surface $\Omega$ and the intrinsic tangential twisting of the normal-aligned octahedral field $f(r)$. The values of $H$ and $K$ simply contribute a fixed quantity depending on $\Omega$ rather than the field. Therefore, the influence of $f$ on $\|\nabla f\|^2_F$ is just in $w$, the intrinsic twist of the cross field it represents.  We stress that this behavior is quite different from the behavior of the component-wise derivative evaluated on vectors, as studied in~\cite{huang_extrinsically_2016}, where their smoothness energy promotes alignment to extrinsic curvature directions.

\subsection{$L^p$ Smoothness Energy of SH Cross Fields}

Suppose we wish to measure smoothness of a normal-aligned octahedral field in the SH representation. We define the following class of convex smoothness energies 
using the $L^p$-norm of $\|\nabla f\|_F$ over the surface $\Omega$ for $p\geq1$:

\begin{equation}
\begin{aligned}
    E_p(\Omega,f) = \left(\int_{\Omega} \|\nabla f\|^p_F~dA\right)^{\frac{1}{p}}.
\end{aligned}\label{eq:LpEnergy}
\end{equation}
We now analyze the behavior of the $E_p$ energy for cross fields in several select cases.

\subsubsection{Case $p=2$: Dirichlet Energy}
We begin with a common choice in geometry processing when smoothness is desirable: the Dirichlet energy $E_2$. Given Proposition \ref{prop:txty}, we can write the Dirichlet energy as $\int_{\Omega} 2H^2 - K + w$. 
Since $H$ and $K$ are independent of the octahedral field $f$, they have no influence over the $f$ that minimizes $E_2$.  Therefore on smooth $\Omega$, we recover intrinsically smooth cross fields. 

Since the Dirichlet energy may diverge at singularities \cite{knoppel_globally_2013}, this choice of energy has the theoretical drawback of diverging for all $f$ in the neighborhood of creases which break octahedral symmetry. In the discretized setting, however, the behavior of $E_2$ is dependent on mesh resolution and empirically leads to strong feature alignment as demonstrated in \S\ref{sec:results}. 
It also leads to an easily-solved optimization problem described in~\S \ref{sec:L2Discretize}.

\subsubsection{Case $p=1$: Vectorial Total Variation}

As noted in the previous section, the conventional means of measuring field smoothness fails to be well-defined for our field representation on creased surfaces. We show here that the $E_1$ energy is not only finite across sharp edges and around singular points but also provides an intuitive measure of field quality that captures both smoothness and feature alignment. It is also known as the \emph{vectorial total variation}.

Consider a function $f:\Omega\rightarrow \R^9$ that is piecewise smooth on $n$ closed patches $\Omega_j$ intersecting in a finite-length curve network $\Gamma = \bigcup_{k=1}^s \gamma_k$, where each $\gamma_k$ is a $C^1$ curve. Equivalently $\Gamma = \bigcup_{j=1}^n \partial \Omega_j$, and the vectorial total variation can be decomposed into integrals over each patch and $\Gamma$. 

\begin{proposition}
	\label{prop:tvfinite}
	For compact $\Omega$ and $f$ as above, $VTV[f]$ is finite and given by the following equation: 
	\begin{equation}
	\label{eq:VTVformula}
    \begin{aligned}
    VTV[f] = \sum_{j=1}^n \int_{\mathring{\Omega}_j} \|\nabla f\|_F~dA + \sum_{k=1}^s \int_{\gamma_k} \|f^+ - f^-\|_2~dL,
    \end{aligned}
    \end{equation}
    where $f^+$ and $f^-$ refer to the limiting values of $f$ on either side of $\gamma_k$, and $\mathring\Omega_j$ denotes the interior of $\Omega_j$.
\end{proposition}
The basic argument starts from Equation $\eqref{eq:vtv}$, splits it into integrals over the patches, applies integration by parts, and utilizes partitions of unity to construct a maximizing sequence of $\phi$'s. The full argument is contained in supplementary materials \S 2. An analogous result, which applies to arbitrary functions on $\R^n$ with bounded variation, is contained in \citep{ambrosio_functions_2000}, with the addition of a third term representing the \emph{Cantor} part of $f$. Since our $f$ is piecewise smooth, however, we can safely ignore the Cantor part. The second term is often referred to as the \emph{jump} part in the total variation literature.

The formula~\eqref{eq:VTVformula} provides an intuitive description of the total variation of an octahedral field in the SH basis as a measure of intrinsic smoothness with extra jump terms. Letting $f$ represent a normal-aligned octahedral field we obtain:
\begin{equation}
\begin{aligned}\label{eq:vtvExpanded}
    VTV[f] &= \sum_{j=1}^n \int_{\mathring{\Omega}_j} \sqrt{2H^2-K+w}~dA + \sum_{k=1}^s \int_{\gamma_k} \left\|f^+ - f^-\right\|_2 ~dL
\end{aligned}
\end{equation}

\paragraph{Generalizing to Creased Surfaces}
While the above result is derived for smooth surfaces $\Omega$ and discontinuous $f$, we can further generalize the result to a surface $\Omega$ constructed from smooth open patches $\Omega_j$ joined along a network of sharp creases $\Gamma = \bigcup_{k=1}^s \gamma_k$. As there is neither a consistent metric nor a consistent tangent space on $\Omega$ across $\Gamma$, 
there is no well-defined choice of gradient. We therefore use equation \eqref{eq:vtvExpanded} as the \emph{definition} for $E_1$ on such a creased surface. Since $f$ is a normal-aligned octahedral field, it is necessarily discontinuous across creases, resulting in contributions to the jump term.

The jump $\|f^+ - f^-\|_2$, where $f^+$ and $f^-$ represent octahedral frames aligned to different normal directions, is minimized if $f^+$ and $f^-$ are both aligned to the axis of rotation from one normal to the other. We formalize this property by Proposition~\ref{prop:featureAlignment}

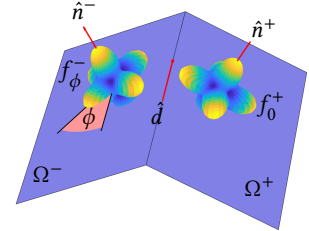
\begin{wrapfigure}[9]{r}{0.22\textwidth}
  \vspace{-14pt}
  \hbox{
  \hspace{-15pt}
  \centering
    \includegraphics[width=0.22\textwidth,draft=false]{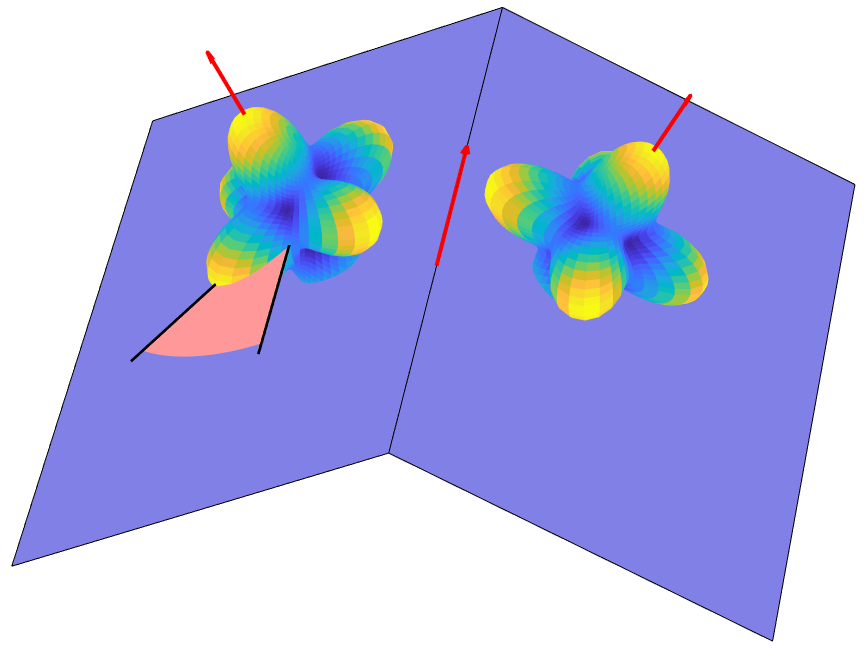}
    \put(-95,60){$f_{\phi}^-$}
    \put(-90,82){$\hat{n}^-$}
    \put(-22,75){$\hat{n}^+$}
    \put(-60,42){$\hat{d}$}
    \put(-20,47){$f_0^+$}
    \put(-86,42){$\phi$}
    \put(-105,20){$\Omega^-$}
    \put(-25,15){$\Omega^+$}
    }
  \vspace{-13pt}
  \captionsetup{justification=raggedright,labelformat=empty}
  \caption{Octahedral frames near crease}
\end{wrapfigure}

\begin{proposition}
    Let $\Omega^+$ and $\Omega^-$ be smooth patches of a surface with normal directions $\hat{n}^+$ and $\hat{n}^-$ that meet at a crease. Let $\hat{d}$ denote the intersection of their tangent spaces at the crease. Let $f_{\theta}^+$ and $f_{\phi}^-$ be the octahedral frames on either side of the crease aligned to $\hat{n}^+$ and $\hat{n}^-$ respectively. $\theta$ and $\phi$ denote their deviation from alignment to $\hat{d}$. The cost $\|f_{\theta}^+-f_{\phi}^-\|_2$ is minimized by $\theta=\phi=0$.
     \label{prop:featureAlignment}
\end{proposition}

Proof of this proposition is left to supplementary materials \S 3.

     The setup is depicted on the right, showing discontinuous normal directions $\hat{n}^+$ and $\hat{n}^-$ as the left and right red arrows respectively. The crease direction $\hat{d}$ is shown by the middle red arrow.  
We emphasize that this proposition implies (locally) crease alignment always minimizes the VTV.  We extensively test and show in supplemental materials that this crease alignment tends to globally hold on surfaces with complicated geometry and topology as well.

\subsubsection{General $p\geq2$}
By Equation~\ref{eq:LpEnergy}, and Proposition~\ref{prop:txty}, $E_p$ incentivizes intrinsic smoothness for all $p$ on smooth domains. On creased domains, 
we have demonstrated (local) crease-alignment for the $p=1$ case. 
For $p\geq2$, the value of $E_p$ diverges for a creased surface. However, we find empirically that minimizing $E_p$ (by recovering solutions to Equation~\ref{eq:smooth_frames_problem}) on a discretized surface leads to stronger feature alignment as $p$ increases.
This behavior may be explained by Proposition~\ref{prop:featureAlignment}, which affects all edges regardless of $p$. The $p$ simply exponentiates the energy across each edge before accumulating it into the total $E_p$. Local to a single edge, the energy-minimizing configuration is unaffected by $p$. Based on our experiments, we further conjecture that the sequence of fields obtained by minimizing $E_2$ on an increasingly dense discrete approximation of $\Omega$ converges to a feature-aligned cross field. This intrinsically smooth feature-alignment is empirically shown in Figure \ref{fig:folded_cylinders}. We leave proof of this conjecture to future work.

\subsubsection{Relation to Polycube Surfaces}\label{sec:sphrepcost}
We achieve an additional property for all values of $p$
through our use of SH octahedral frames. Consider the case of $\Omega$ being a cube: minimizers of $E_p$ will have zero energy, despite the cube's sharp corners, since the field's octahedral symmetry allows it to simultaneously align to all three creases at each corner.
Effectively a surface with many angle-$\frac{\pi}{2}$ turns and cube-corners can have just as low of an energy as one with no creases at all. More generally, if $\Omega$ is a polycube surface, $E_p(\Omega,f)=0$ by choosing $f$ to be a facet aligned uniform frame field. 

\section{Optimizing for an Octahedral Frame Field}

Our discussion above provides a new class of energies based on the SH representation of cross fields, which naturally promote both intrinsic smoothness and extrinsic crease alignment without the need for feature curve detection or reliance on potentially noisy local curvature estimates. For this reason, we propose solving the following variational problem to find a cross field $f^*$ over a surface:
\begin{equation}\label{eq:smooth_frames_problem}
\begin{aligned}
f^* & = \underset{f}{\arg\min}
& & E_p(\Omega,f)\\
& \text{subject to}
& & W_{n(x)} f(x) = u_0.
\end{aligned}
\end{equation}
Recall that the constraint encodes normal alignment of the frame field (see equation \eqref{eq:normalAlignment}).
 Some past algorithms have an extra $||f(x)||_2=1$ constraint which
 results in uniform-scale isotropic octahedral fields over $\Omega$. This constraint makes the problem nonconvex, and causes the functional to diverge in the neighborhood of field singularities, which are unavoidable on generic surfaces by the Poincar\'e-Hopf Theorem.  Accordingly, a relaxation is naturally required; we drop the constraint $\|f(x)\|_2=1$, yielding a convex problem with globally-optimal solution. Dropping the $\|f(x)\|_2=1$ constraint allows the frame's two  tangential components to scale independently from its normal component, resulting in anisotropic octahedral fields. We obtain octahedral fields with uniform-magnitude normal-lobes, and varying scale in the magnitude of the tangential cross field. This relaxation is similar in spirit to the one which appears in
 \citep{knoppel_globally_2013}
 , and has similar benefits, including automatic placement of singularities, and bounded-energy minimizers in the smooth limit (which is necessary in order for the field to be insensitive to the underlying mesh).

\subsection{Soft Normal Alignment}
It is sometimes beneficial to relax the normal alignment constraint, e.g., in cases where the mesh contains sliver triangles with unstable normal directions.  In these cases, a smoother cross field can be obtained by deviating slightly from exact normal alignment. This relaxation changes the optimization problem from Equation~\eqref{eq:smooth_frames_problem} into the following:
\begin{equation}\label{eq:smooth_frames_problem_softN}
\begin{aligned}
f^* & = \underset{f}{\text{argmin}}
& & E_p(\Omega,f)\\
& \text{subject to}
& & \|W_{n(x)} f(x) - u_0\|_2 \leq \epsilon.
\end{aligned}
\end{equation}
This problem imposes a point-wise normal alignment constraint with tolerance $\epsilon$. When $\epsilon=0$, we recover the hard normal alignment formulation \eqref{eq:smooth_frames_problem}. On the opposite side of the spectrum, as $\epsilon \rightarrow \|u_0\|_2 = \sqrt{\frac{7}{12}} \approx\; 0.76$, the solution to \eqref{eq:smooth_frames_problem_softN} approaches a constant octahedral field. This is the case where normal alignment has relaxed so far that the octahedral frames are effectively unconstrained. 
\begin{wrapfigure}[12]{r}{0.30\textwidth}
    \vspace{-10pt}
    \hbox{
    \hspace{-5pt}
    \centering
    \includegraphics[width=0.31\textwidth]{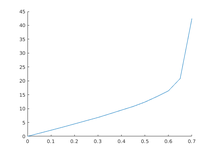}
    \put(-125,-10){Soft normal alignment: $\epsilon$}
    \put(-165,10){\rotatebox{90}{Maximum deviation degree}}
    }
    \captionsetup{labelformat=empty}
    \label{fig:normalDeviation}
\end{wrapfigure}

For values in between we perform the following experiment to obtain a rough correspondence between soft normal alignment parameter $\epsilon$ and maximum angle deviation from normal alignment: For each value of $\epsilon$ between $0$ and $.7$ (at intervals of $.05$), we sample $100000$ $\epsilon$-perturbations of a $\hat{z}-$aligned frame, extract the frame they represent, and compute its maximum angle deviation from the $\hat{z}$-axis. Results are shown in the inset.
 
 We highlight that this parameter encodes a point-wise constraint uniformly applied over the mesh. As such its interpretation does not change with different meshes. Please see the supplemental materials for results on over 200 different meshes using a variety of values of $\epsilon$.
 
\begin{figure}
    \centering
    \begin{subfigure}[t]{0.32\columnwidth}
	\includegraphics[width=\textwidth]{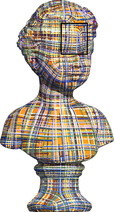}
	\caption{hard normal-aligned streamlines}
	\end{subfigure}%
	\hfill%
	\begin{subfigure}[t]{0.32\columnwidth}
	\includegraphics[width=\textwidth]{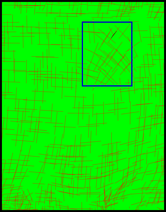}
	\caption{hard normal-aligned crosses}
	\end{subfigure}
	\hfill%
	\begin{subfigure}[t]{0.32\columnwidth}
	\includegraphics[width=\textwidth]{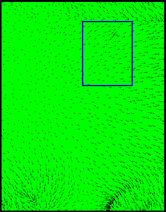}
	\caption{mesh normal directions near sliver triangle (zoom $\times$ 1)}
	\end{subfigure}%
	\\
	
	\begin{subfigure}[t]{0.32\columnwidth}
	\includegraphics[width=\textwidth]{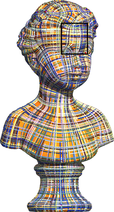}
	\caption{soft normal-aligned streamlines}
	\end{subfigure}%
	\hfill%
	\begin{subfigure}[t]{0.32\columnwidth}
	\includegraphics[width=\textwidth]{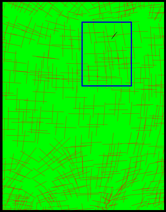}
	\caption{soft normal-aligned crosses}
	\end{subfigure}
	\hfill%
	\begin{subfigure}[t]{0.32\columnwidth}
	\includegraphics[width=\textwidth]{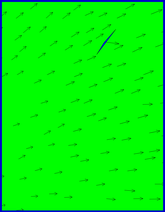}
	\caption{mesh normal directions near sliver triangle (zoom $\times$ 2)}
	\end{subfigure}%
    \caption{Soft normal alignment increases quality of the cross field and decreases the influence of mesh artifacts. The \textbf{buste} mesh is shown with $p=\infty$ and varying normal alignment: $\epsilon = 0$ for the top figure and $\epsilon = 0.5$ on the bottom.
    (a) Hard normal aligned streamlines;
    (b) magnified crosses shows a small patch of diagonal crosses in an otherwise regular region;
    (c) magnified triangle normals visualized with sliver triangle 4611 shaded in blue;
    (d) soft normal aligned streamlines;
    (e) magnified crosses no longer shows diagonal artifacts;
    (f) extra-magnified triangle normals visualized with sliver triangle 4611 shaded in blue. While the normal direction of the region points diagonally up and right, the sliver triangle's normal direction points almost completely to the right.
    }
    \label{fig:softnormalalignment}
\end{figure}

 The benefit of soft normal alignment is demonstrated in Figure~\ref{fig:softnormalalignment}. Due to the influence of a sliver triangle in the \textbf{buste} mesh with unstable normal direction, the hard-normal-aligned cross field is forced to create a localized artifact. By using soft normal alignment, the sliver triangle's unstable normal direction has less influence over the resulting cross field, therefore increasing the quality of the result. A similar benefit is demonstrated on the \textbf{duck} and \textbf{armchair} meshes shown in the supplementary materials.
 
 Additionally we test soft normal alignment on a \textbf{cube-mesh} with artificial noise added in Figure \ref{fig:noisy_brick}. With hard normal alignment the cross fields exhibit undesirable alignment to noise which increases with $p$. With soft-normal alignment, the cross fields show significantly decreased sensitivity to noise.

\subsection{Discretization}
\label{sec:L2Discretize}

Now we describe how to construct smooth cross-fields by numerically optimizing a discretization of $E_p$. 
We assume the surface $\Omega$ has been triangulated into a manifold mesh $\mathcal{M}=\left(V,E,F\right)$. Let $n_t$ be the normal direction of triangle $t\in F$. We represent a cross on $M$ as a normally-aligned octahedral frame $f_t\in\mathcal{V}\subset \R^9$ per triangle. We use the shorthand $f$ to denote the concatenation of all $f_t$ into a single $9|F| \times 1$ vector. $V$ is a $n_v \times 3$ matrix of vertex positions, where $n_v$ is the number of vertices. $E$ denotes the $n_e \times 2$ matrix of edges, where $n_e$ is the number of edges. The energy $E_p$ can be discretized as
\begin{equation}
\label{eq:discreteEp}
E_p = \left(\sum_{e\in E} w_e \|f_{t_1} - f_{t_2}\|_2^p\right)^{\frac{1}{p}}
\end{equation}
where $t_1$ and $t_2$ are triangles adjacent to edge $e$, and $w_e$ are weights corresponding to the dual Laplacian. We use $w_e=\frac{\|e\|}{\|e^*\|}$, where $\|e\|$ is the length of edge $e$ and $\|e^*\|$ is the distance between barycenters of $t_1$ and $t_2$. 

For $\epsilon=0$, the normal alignment constraint is discretized by the  linear constraint $W f = u$, where $W$ is a sparse block-diagonal matrix with a block $W_{n_t}$ for each triangle. It has dimensions $7|F|\times9|F|$. The vector $u$ is a repetition of $u_0$ for each triangle, resulting in a $7|F|\times1$ vector. For $\epsilon>0$, the normal alignment constraint is discretized by a second-order cone constraint: $\|W_{n_t}f_t-u_0\|_2\leq \epsilon$ per triangle.

For the case $p=1$ and a completely flat surface $\Omega$, our discretization agrees with the standard discretization of total variation in image processing \cite{rudin_nonlinear_1992,chambolle_introduction_2010}.

\subsection{Meshes with Boundary}\label{sec:boundary_case}
When the mesh contains boundaries, we do not enforce any boundary conditions; in PDE parlance, this choice corresponds to ``natural boundary conditions.'' While it is tempting to expect the natural boundary conditions for $p=2$ to imply zero Neumann boundary conditions \cite{stein_natural_2018}, the SH representation vector is complicated by being constrained to a spatially varying linear subspace.  
We simply allow the cross on the boundary to be that which minimizes total energy. If desired, one can enforce a %
constraint that the cross field on the boundary be aligned to the boundary through the method described in \S\ref{sec:manual_guidance}.

\subsection{Manual Guidance}\label{sec:manual_guidance}
\begin{figure}
    \hfill%
    \begin{subfigure}[t]{0.44\columnwidth}
	\includegraphics[width=1\textwidth]{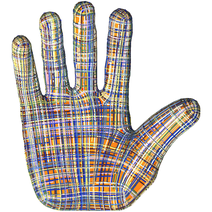}
	\caption{before}
	\end{subfigure}%
	\hfill%
	\begin{subfigure}[t]{0.48\columnwidth}
	\includegraphics[width=1\textwidth]{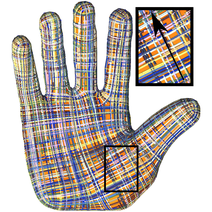}
	\caption{after}
	\hfill%
	\vspace{-10pt}
	\end{subfigure}
	\caption{Octahedral fields obtained by minimizing $E_{\infty}$ on the \textbf{hand} mesh before and after adding manual direction. The manually-added streamline is shown by the inset black arrow. This constraint removes a singularity from the original octahedral field. 
	} 
	\label{fig:manualAlignmentConstraintFig}
\end{figure}

\begin{figure}[t]
    \captionsetup[subfigure]{labelformat=empty}
	\begin{subfigure}[t]{0.24\columnwidth}
	\includegraphics[width=1\textwidth]{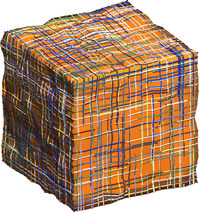}
	\caption{$p=2,\epsilon=0$}
	\end{subfigure}
	\hfill
	\begin{subfigure}[t]{0.24\columnwidth}
	\includegraphics[width=1\textwidth]{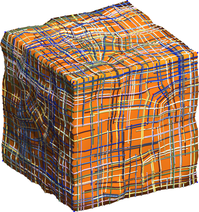}
	\caption{$p=3,\epsilon=0$}
	\end{subfigure}
	\hfill
	\begin{subfigure}[t]{0.24\columnwidth}
	\includegraphics[width=1\textwidth]{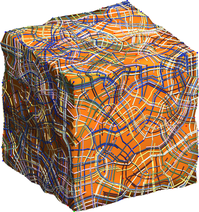}
	\caption{$p=6,\epsilon=0$}
	\end{subfigure}
	\hfill
	\begin{subfigure}[t]{0.24\columnwidth}
	\includegraphics[width=1\textwidth]{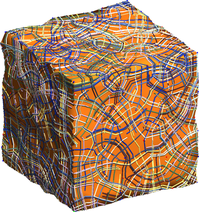}
	\caption{$p=8,\epsilon=0$}
	\end{subfigure}
	\hfill
	\\
	\begin{subfigure}[t]{0.24\columnwidth}
	\includegraphics[width=1\textwidth]{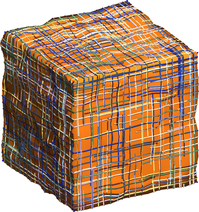}
	\caption{$p=2,\epsilon=.2$}
	\end{subfigure}
	\hfill
	\begin{subfigure}[t]{0.24\columnwidth}
	\includegraphics[width=1\textwidth]{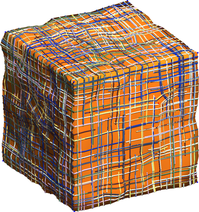}
	\caption{$p=3,\epsilon=.2$}
	\end{subfigure}
	\hfill
	\begin{subfigure}[t]{0.24\columnwidth}
	\includegraphics[width=1\textwidth]{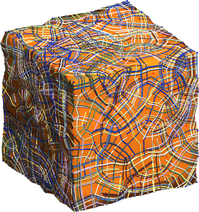}
	\caption{$p=6,\epsilon=.2$}
	\end{subfigure}
	\hfill
	\begin{subfigure}[t]{0.24\columnwidth}
	\includegraphics[width=1\textwidth]{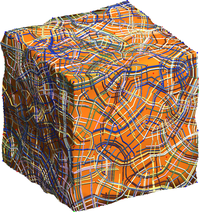}
	\caption{$p=8,\epsilon=.2$}
	\end{subfigure}
	\hfill
	\\
	\begin{subfigure}[t]{0.24\columnwidth}
	\includegraphics[width=1\textwidth]{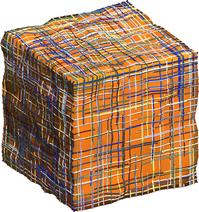}
	\caption{$p=2,\epsilon=.525$}
	\end{subfigure}
	\hfill
	\begin{subfigure}[t]{0.24\columnwidth}
	\includegraphics[width=1\textwidth]{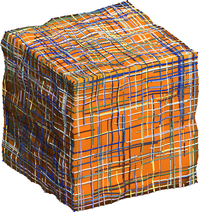}
	\caption{$p=3,\epsilon=.525$}
	\end{subfigure}
	\hfill
	\begin{subfigure}[t]{0.24\columnwidth}
	\includegraphics[width=1\textwidth]{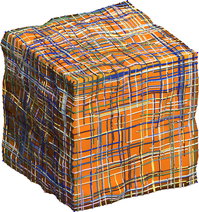}
	\caption{$p=6,\epsilon=.525$}
	\end{subfigure}
	\hfill
	\begin{subfigure}[t]{0.24\columnwidth}
	\includegraphics[width=1\textwidth]{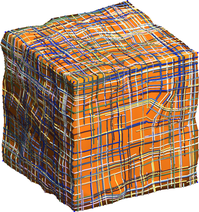}
	\caption{$p=8,\epsilon=.525$}
	\end{subfigure}
	\hfill
	\\
	\begin{subfigure}[t]{0.24\columnwidth}
	\includegraphics[width=1\textwidth]{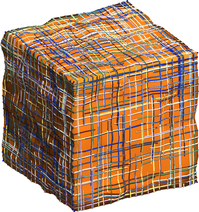}
	\caption{$p=2,\epsilon=.55$}
	\end{subfigure}
	\hfill
	\begin{subfigure}[t]{0.24\columnwidth}
	\includegraphics[width=1\textwidth]{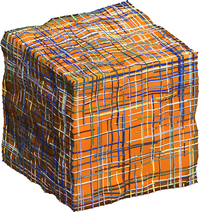}
	\caption{$p=3,\epsilon=.55$}
	\end{subfigure}
	\hfill
	\begin{subfigure}[t]{0.24\columnwidth}
	\includegraphics[width=1\textwidth]{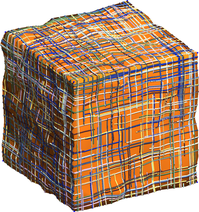}
	\caption{$p=6,\epsilon=.55$}
	\end{subfigure}
	\hfill
	\begin{subfigure}[t]{0.24\columnwidth}
	\includegraphics[width=1\textwidth]{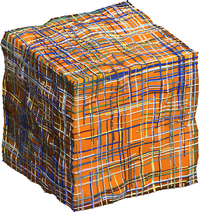}
	\caption{$p=8,\epsilon=.55$}
	\end{subfigure}
	\hfill
	\caption{As $\epsilon$ increases or as $p$ decreases, the cross fields become less sensitive to noise added to the \textbf{cube-mesh}. }
	\label{fig:noisy_brick} 
\end{figure}

To support manual guidance of the octahedral frame field, we can prescribe alignment of the frame field to streamlines. Streamline constraints combined with normal alignment result in a fully-determined frame. Therefore, prescribing streamlines is equivalent to prescribing the value of $f_t$ on a subset of triangles $T_p$. Denote the prescribed octahedral frame on triangle $t$ as $F_t$. We then add a new linear constraint that 
\begin{equation}\label{eq:prescribe_streamlines}
    \forall t \in T_p, \;\; f_t = F_t.
\end{equation}
This technique is demonstrated in Figure~\ref{fig:manualAlignmentConstraintFig}.

\subsection{Non-Triviality Constraint} \label{sec:nontriviality}
 As a result of dropping the unit-norm constraint from Equation~\ref{eq:smooth_frames_problem}, we have no explicit guarantee that the tangential components of octahedral frames do not degenerate to zero. 
On a surface with a crease, however, the normal alignment constraint on one side of the crease imposes that the magnitude of the tangential component on the other side of the crease is close to one. As a result, we observe empirically that the vast majority of our octahedral frames do not degenerate. 

In the case that octahedral frames do degenerate significantly, their norms can be too small to project robustly. We locate these by using the octahedral projection from \citep{anonymous_algebraic_2019} to measure  the distance from $f_t$ to the octahedral variety $\mathcal{V}$ : $d(f_t) = \|\pi_{\mathcal{V}}(f_t)-f_t\|_2$, and thresholding by $d(f_t) > .665$. If such frames are found, we run the optimization again while holding non-degenerate frames to their projected values. In our experiments, just one round of re-solving results in 99.8\% non-degenerate frames. 

\subsection{Solving For an Octahedral Field}
In its most general form, our problem formulation consists of minimizing a mixed-norm objective, with both linear and  second-order cone constraints. This results in a convex problem that we solve with Mosek 9
~\cite{aps_mosek_2017}. 
The normal alignment constraint becomes $\left[ \epsilon, (W_{n_t}f_t-u_0)^T \right] \in \mathcal{L}^8$, where $\mathcal{L}^8$ is the 8-dimensional Lorentz cone. Likewise, the energy is formulated using a single $p$-norm cone.
Our code is written in Matlab with a \texttt{mex} interface to Mosek; it builds cross-platform. Since our problem is convex, any dependence on initialization would entirely be due to non-unique solutions, which we do not observe in practice. Furthermore, we use the interior point method, which does not accept manual initialization. In the specific case of $\epsilon=0,p=2$, solving this optimization is equivalent to solving a linear system.

\section{Results}\label{sec:results}

\begin{figure}[t]
    \captionsetup[subfigure]{labelformat=empty}
	\rotatebox{90}{\;\;\;\;\;$p=2$}
	\begin{subfigure}[t]{0.23\columnwidth}
	\includegraphics[width=1\textwidth]{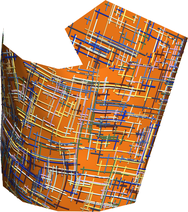}
	\end{subfigure}
	\hfill
	\begin{subfigure}[t]{0.23\columnwidth}
	\includegraphics[width=1\textwidth]{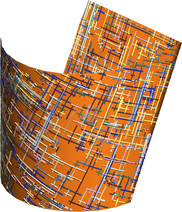}
	\end{subfigure}
	\hfill
	\begin{subfigure}[t]{0.23\columnwidth}
	\includegraphics[width=1\textwidth]{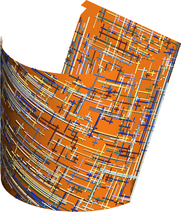}
	\end{subfigure}
	\hfill
	\begin{subfigure}[t]{0.23\columnwidth}
	\includegraphics[width=1\textwidth]{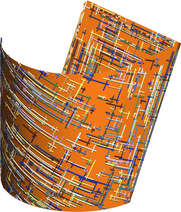}
	\end{subfigure}
	\hfill 
	\\
	\rotatebox{90}{\;\;\;\;\;$p=3$}
	\begin{subfigure}[t]{0.23\columnwidth}
    \includegraphics[width=1\textwidth]{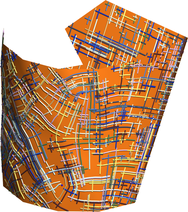}
	\end{subfigure}
	\hfill
	\begin{subfigure}[t]{0.23\columnwidth}
	\includegraphics[width=1\textwidth]{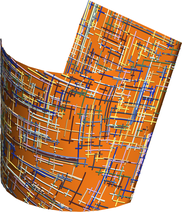}
	\end{subfigure}
	\hfill
	\begin{subfigure}[t]{0.23\columnwidth}
	\includegraphics[width=1\textwidth]{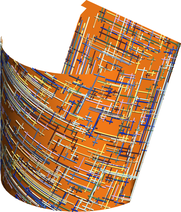}
	\end{subfigure}
	\hfill
	\begin{subfigure}[t]{0.23\columnwidth}
	\includegraphics[width=1\textwidth]{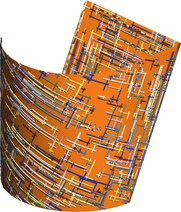}
	\end{subfigure}
	\hfill
	\\
	\rotatebox{90}{\;\;\;\;\;Meshes}
	\begin{subfigure}[t]{0.23\columnwidth}
	\includegraphics[width=1\textwidth]{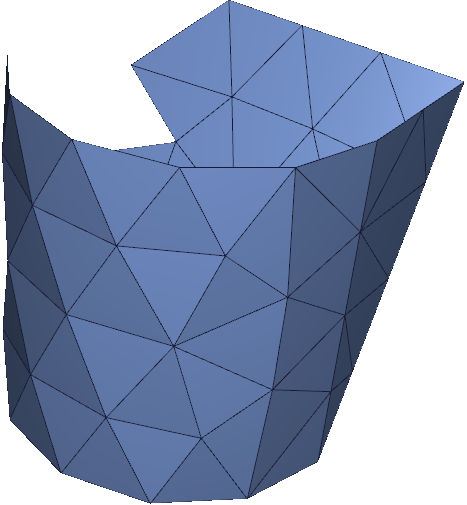}
	\caption{100 Triangles}
	\end{subfigure}
	\hfill
	\begin{subfigure}[t]{0.23\columnwidth}
	\includegraphics[width=1\textwidth]{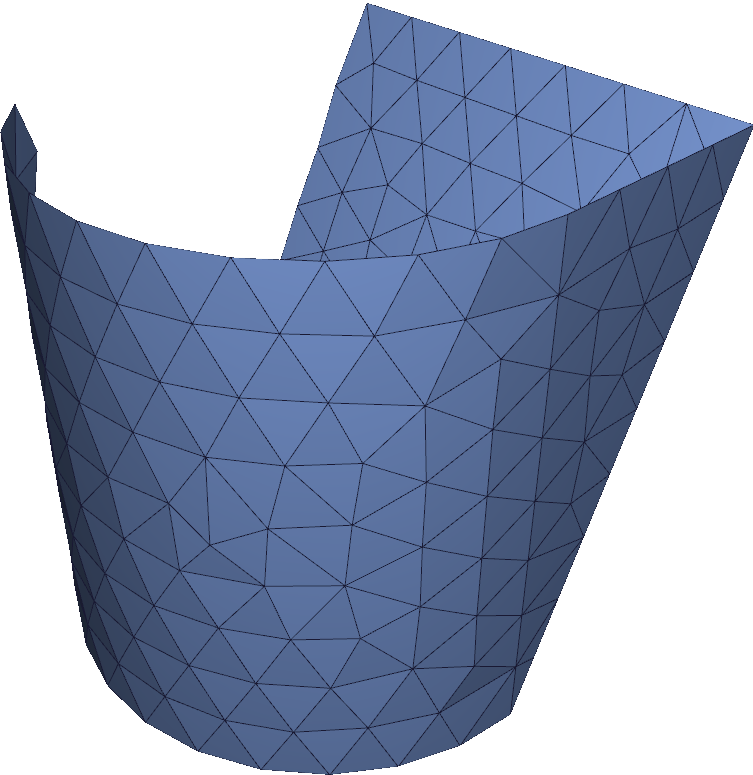}
	\caption{400 Triangles}
	\end{subfigure}
	\hfill
	\begin{subfigure}[t]{0.23\columnwidth}
	\includegraphics[width=1\textwidth]{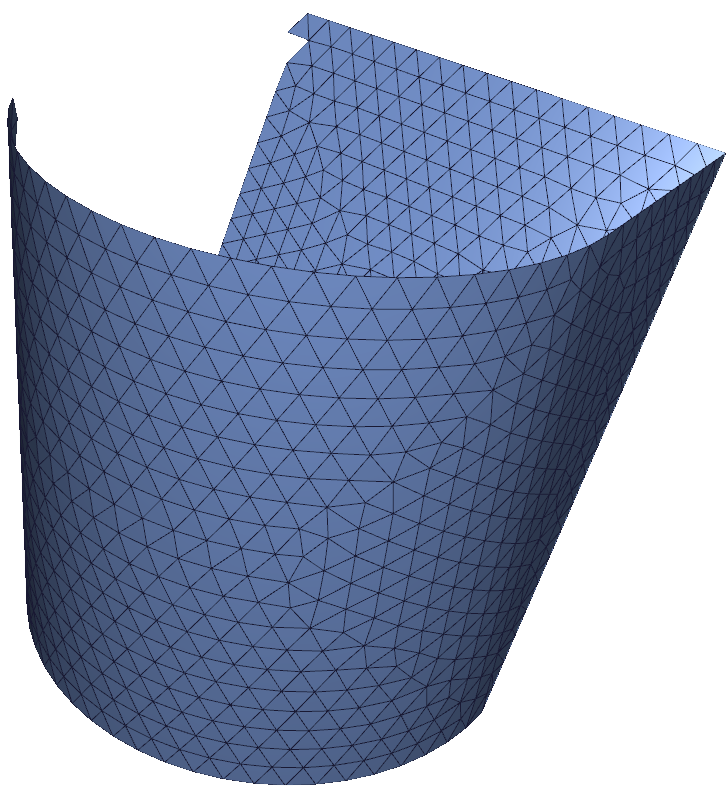}
	\caption{2K Triangles}
	\end{subfigure}
	\hfill
	\begin{subfigure}[t]{0.23\columnwidth}
	\includegraphics[width=1\textwidth]{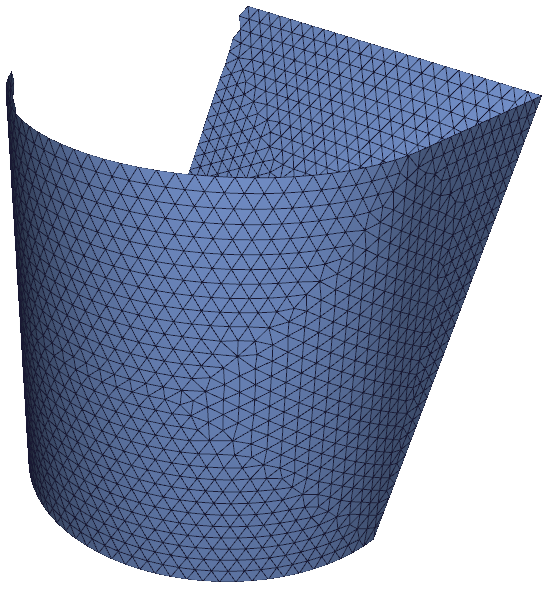}
	\caption{4K Triangles}
	\end{subfigure}
	\hfill 
	\caption{On this developable surface, our cross fields are intrinsically smooth in the limit of refinement, but exhibit some mesh sensitivity on coarse meshes, particularly for higher $p$ values. They are crease-aligned for all resolutions. Note that the extrinsic curvature of the cylindrical bend has no effect on the cross fields at higher resolutions.}
	\label{fig:folded_cylinders} 
\end{figure}

\begin{figure*}[t]
    \begin{subfigure}[t]{0.24\textwidth}
	\includegraphics[width=1\textwidth]{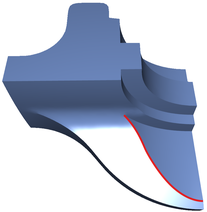}
	\caption{Fandisk mesh}
	\end{subfigure}%
	\hfill%
    \begin{subfigure}[t]{0.24\textwidth}
	\includegraphics[width=1\textwidth]{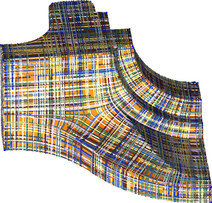}
	\caption{$p=1$}
	\end{subfigure}%
	\hfill%
    \begin{subfigure}[t]{0.24\textwidth}
	\includegraphics[width=1\textwidth]{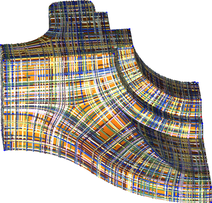}
	\caption{$p=2$}
	\end{subfigure}%
	\hfill%
    \begin{subfigure}[t]{0.24\textwidth}
	\includegraphics[width=1\textwidth]{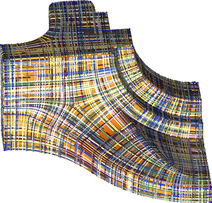}
	\caption{$p=\infty$}
	\end{subfigure}%
	\hfill%
	\caption{Cross fields generated by minimizing $E_p$ for $p=1,2,\infty$ on the \textbf{fandisk} mesh. The shallow crease of the \textbf{fandisk} mesh is marked in red. Our cross fields naturally align to the shallow crease with increasing strength for higher $p$.}
	\label{fig:VaryingWithPFieldsFig}
\end{figure*}

\begin{figure}[t]
    \begin{subfigure}[t]{0.19\columnwidth}
	\includegraphics[width=1\textwidth]{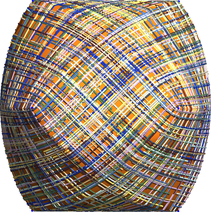}
	\caption{1.3k}
	\end{subfigure}%
	\hfill%
    \begin{subfigure}[t]{0.19\columnwidth}
	\includegraphics[width=1\textwidth]{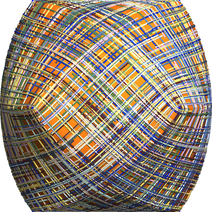}
	\caption{5.4k}
	\end{subfigure}%
	\hfill%
    \begin{subfigure}[t]{0.19\columnwidth}
	\includegraphics[width=1\textwidth]{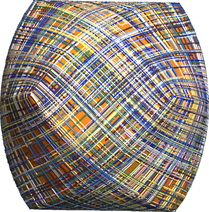}
	\caption{21.8k}
	\end{subfigure}%
	\hfill%
    \begin{subfigure}[t]{0.19\columnwidth}
	\includegraphics[width=1\textwidth]{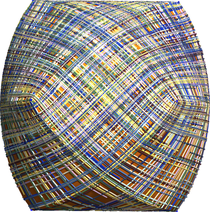}
	\caption{11.6k}
	\end{subfigure}%
	\hfill%
    \begin{subfigure}[t]{0.19\columnwidth}
	\includegraphics[width=1\textwidth]{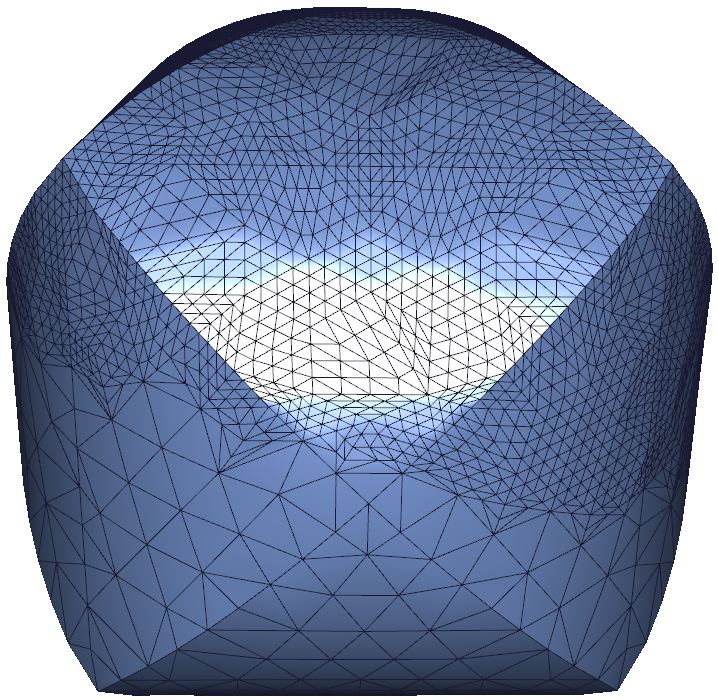}
	\caption{Multires}
	\end{subfigure}%
	\hfill%
	\caption{Cross fields generated by minimizing $E_2$ on different meshings of the \textbf{three-cylinder-intersection} with number of faces indicated. Cross field (d) is computed on the multi-resolution mesh (e). Notice that we obtain the same feature-aligned cross field each time.}
	\label{fig:multires}
\end{figure}

We begin with a comparison of the behavior of our energy for different values of $p$. This experiment is depicted in Figure~\ref{fig:VaryingWithPFieldsFig}. We observe that our cross fields naturally align to features with increasing strength for higher $p$. In the case $p=1$, our cross field is discontinuous over all creases, but while it is provably incentivized to align, it sometimes deviates due to the influence of neighboring creases, e.g. on the top surface of the fandisk. For $p=2$, our cross fields achieve close alignment to the upper half of the shallow crease, as well as alignment on the top face where the $p=1$ case failed. Finally for $p=\infty$ our fields align down the entirety of the shallow crease. While in theory, $E_p$ for $p\geq2$ diverges on creases, we observe that its discretization yields empirically strong crease alignment. This may be due to application of Proposition~\ref{prop:featureAlignment} over all edges of the mesh. 
We show our fields for different discretizations of the same geometry in Figure~\ref{fig:multires} and observe that in all cases, we achieve crease alignment. 

\paragraph{Supplementary Materials} In our supplementary document we perform an empirical study to evaluate the performance of our method.  We evaluate our method on a number of models drawn from the Thingi10k~\cite{Thingi10K} dataset, as well as a number of other commonly used benchmark models to demonstrate effective crease alignment on real-world models.  We also compare our approach to several baseline methods (\cite{jakob_instant_2015,brandt_modeling_2018,knoppel_globally_2013}), by generating fields on the models in the ``Robust Field-Aligned Global Parametrization'' dataset~\cite{Myles:RFAGP}, taking care to sample the relevant parameter space for each formulation.  
While it is difficult to precisely quantify the quality of a vector field, we highlight a number of cases where our method recovers fields which more faithfully conform to mesh features than baseline methods on real-world models.

Our runtimes are shown for a 
set of meshes with 240 to 76K vertices and 480 to 152K faces in Figure~\ref{fig:runtimes}. Runtimes naturally increase with mesh size and appear to grow linearly with number of triangles in our mesh test set. Memory costs are incurred to store a $W_{n_t}$ per triangle, a single $u_0$, $w_e$ per edge, and $f_t$ per triangle. Hence, storage is linear in size of the mesh. More detailed information regarding parameter choices and runtimes is provided in the supplementary materials. Table \ref{table:timings} shows a summary of our runtimes in comparison to that of other methods. Our runtimes are on the same scale as
\citep{knoppel_globally_2013} and to the bases setup step in \citep{brandt_modeling_2018}.
\begin{table}[]
\scalebox{.9}{
\begin{tabular}{| r | rr | r | r | r |}
\hline
Number of & Bases    & Biharmonic & Instant     & Globally    & Ours \\ 
triangles & setup    & solve      & meshes       & optimal     & \\ \hline
3K    & 5   & .005 & .026     & .85    & 2.8    \\ \hline
12K   & 24  & .005 & .053     & 20.46  & 15.058 \\ \hline
20K   & 44  & .005 & .080     & 21.913 & 25.895 \\ \hline
69K   & 170 & .006 & .141     & 62.733 & 135.09 \\ \hline
80K   & 181 & .006 & .222     & 71.15  & 112.3  \\ \hline
\end{tabular}}
\caption{Runtimes in seconds for computing cross-fields using different methods on meshes with a varying number of triangles. Methods listed are those of  \citet{brandt_modeling_2018}, \citet{jakob_instant_2015}, \citet{knoppel_globally_2013}, and our own. Runtimes for fields from
\cite{brandt_modeling_2018}
are split into time needed for the setup of 500 bases eigenfields and the field computation separately because of drastically differing timescales.}
\label{table:timings}
\end{table}

\begin{figure}[t]
    \begin{subfigure}[t]{0.3\columnwidth}
	\includegraphics[width=1\textwidth]{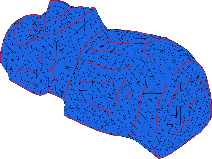}
	\caption{\textbf{Moai} and explicit feature curves(red edges)}
	\end{subfigure}%
	\hfill%
    \begin{subfigure}[t]{0.3\columnwidth}
	\includegraphics[width=1\textwidth]{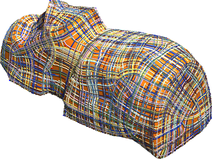}
	\caption{Explicit Feature Curve Alignment: $p=2, \epsilon=0$}
	\end{subfigure}%
	\hfill%
    \begin{subfigure}[t]{0.3\columnwidth}
	\includegraphics[width=1\textwidth]{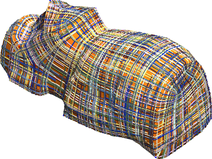}
	\caption{Our method: $p=2$, $\epsilon=0$}
	\end{subfigure}%
	\hfill%
	\caption{Comparison of our feature-aligned cross fields to those generated when adding additional explicit feature curve alignment constraints. Explicit feature curves were obtained from
	\citep[Fig. 9]{gehre_adapting_2016}
	Despite the extra cost of precomputing explicit feature curves, slight artifacts in the feature curves (most pronounced on the side) force the explicitly guided cross field to have lower quality.
    }
	\label{fig:sparsefeatureconstraints}
\end{figure}

\paragraph{Comparison to Explicit Feature Curves.} Next, we compare our feature-aligned cross fields to those produced with the help of explicitly-computed feature curves. We obtain feature curves on the 1904-triangle \textbf{Moai} mesh from 
\citep[Fig. 9]{gehre_adapting_2016}. 
We compute a cross field with additional hard constraints as described in \S \ref{sec:manual_guidance} to enforce alignment to the precomputed feature curves. We compare the resulting field with and without explicit feature-curve alignment in Figure~ \ref{fig:sparsefeatureconstraints}. While the feature curves help guide the cross field, just a few artifacts in the computed features drastically influence the resulting cross field to have more singularities and be less smooth without clear benefit. The \textbf{Moai} is shown from an angle where these differences are most pronounced.

\paragraph{Effect of Mesh Resolution on Crease-Alignment vs Extrinsic Curvature.} In this experiment we test on a geometry where a sharp crease is mis-aligned to extrinsic curvature directions. We generate meshes of this geometry at varying resolution to see how crease-alignment interacts with extrinsic curvature. Results of this experiment are depicted in Figure \ref{fig:folded_cylinders}. As mesh resolution increases our cross fields become crease-aligned and intrinsically smooth, agreeing with the theory. For very low mesh resolution, the cross fields are more sensitive to the underlying meshing pattern.

\begin{figure}[t]
    \begin{subfigure}[t]{0.3\columnwidth}
	\includegraphics[width=1\textwidth,angle=180]{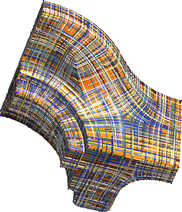}
	\caption{Ours $p=2$, $\epsilon=0$: 3.9s }
	\end{subfigure}%
	\hfill%
    \begin{subfigure}[t]{0.3\columnwidth}
	\includegraphics[width=1\textwidth,angle=180]{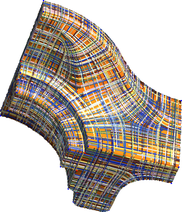}
	\caption{Ours $p=\infty$, $\epsilon=0$: 3.5s}
	\end{subfigure}%
	\hfill%
    \begin{subfigure}[t]{0.3\columnwidth}
	\includegraphics[width=1\textwidth,angle=180]{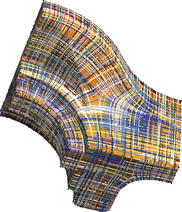}
	\caption{BEM: 161s}
	\end{subfigure}%
	\hfill%
	\caption{Cross field and runtime comparison of our method to a method optimizing volumetric octahedral frames~\cite{solomon_boundary_2017}. The \textbf{fandisk} used contains 2.5k triangles.}
	\label{fig:BEMcomparison}
\end{figure}

\begin{figure}
    \centering
    \includegraphics[scale=.7]{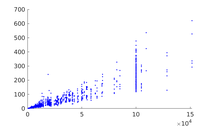}
    \put(-150,0){Number of triangles}
    \put(-250,50){\rotatebox{90}{Time in seconds}}
    \caption{Runtimes to compute cross fields over various mesh sizes.}
    \label{fig:runtimes}
\end{figure}

\paragraph{Comparison to 3D Octahedral Fields.} Due to similarity of frame representation, we compare our method to surface cross fields obtained by optimizing a volumetric octahedral field. Algorithms like those of \citep{huang_boundary_2011,ray_practical_2016} can generate surface cross fields by approximating the surface with the limiting behavior of a thin layer of tetrahedra or prism elements. However, prism elements are non-standard and both element types will be poorly conditioned without introducing further restrictions such as zero normal gradient to mimic a triangle mesh. We instead opt to compare with 
 the Boundary Element Method (BEM)~\cite{solomon_boundary_2017} which acts directly on surface triangle meshes. 
We use the 2500 triangle \textbf{fandisk} mesh for this comparison. As observed earlier, our method has increasing feature alignment with increased values of $p$. In comparison, Figure~\ref{fig:BEMcomparison} shows that the BEM field fully ignores the shallow crease of the \textbf{fandisk}, running through it at a $45^{\circ}$ offset. Moreover, despite the fact that the BEM only needs boundary data as input, its runtime is close to 50 times slower than ours.

\begin{figure*}[t]
	\begin{subfigure}[t]{0.15\textwidth}
	\caption{Ours $p=2$ \hfill \\ \qquad \qquad \qquad \qquad  }
	\includegraphics[width=1\textwidth]{figures/resdep/shrunk_Cyl3Inter_denser-streamlines.png}
	\end{subfigure}
	\hfill
	\begin{subfigure}[t]{0.15\textwidth}
	\caption{\cite{jakob_instant_2015}\qquad \qquad}
	\includegraphics[width=1\textwidth]{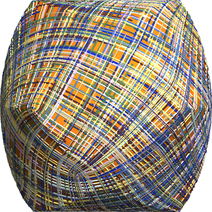}
	\end{subfigure}
	\hfill
	\begin{subfigure}[t]{0.15\textwidth}
	\caption{\cite{brandt_modeling_2018} $\lambda = -.1$  Biharmonic}
	\includegraphics[width=1\textwidth]{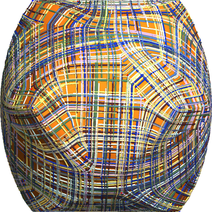}
	\end{subfigure}
	\hfill
	\begin{subfigure}[t]{0.15\textwidth}
	\caption{\cite{brandt_modeling_2018} $\lambda = -.0001$ Biharmonic}
	\includegraphics[width=1\textwidth]{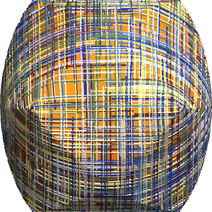}
	\end{subfigure}
	\hfill
	\begin{subfigure}[t]{0.15\textwidth}
	\caption{\cite{knoppel_globally_2013} $\lambda = -.1$ A-holomorphic}
	\includegraphics[width=1\textwidth]{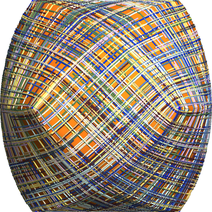}
	\end{subfigure}
	\hfill
	\begin{subfigure}[t]{0.15\textwidth}
	\caption{\cite{knoppel_globally_2013} $\lambda = -.1$ Dirichlet}
	\includegraphics[width=1\textwidth]{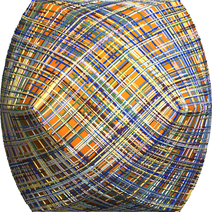}
	\end{subfigure}
	\hfill\\
	\begin{subfigure}[t]{0.15\textwidth}
	\includegraphics[width=1\textwidth]{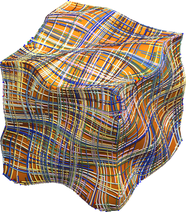}
	\end{subfigure}
	\hfill
	\begin{subfigure}[t]{0.15\textwidth}
	\includegraphics[width=1\textwidth]{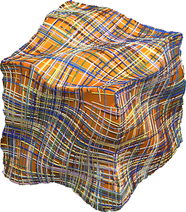}
	\end{subfigure}
	\hfill
	\begin{subfigure}[t]{0.15\textwidth}
	\includegraphics[width=1\textwidth]{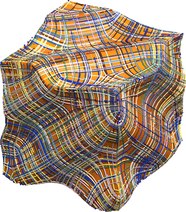}
	\end{subfigure}
	\hfill
	\begin{subfigure}[t]{0.15\textwidth}
	\includegraphics[width=1\textwidth]{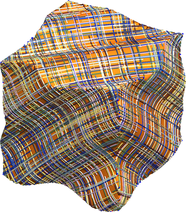}
	\end{subfigure}
	\hfill
	\begin{subfigure}[t]{0.15\textwidth}
	\includegraphics[width=1\textwidth]{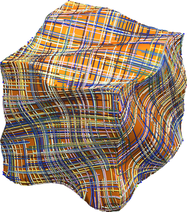}
	\end{subfigure}
	\hfill
	\begin{subfigure}[t]{0.15\textwidth}
	\includegraphics[width=1\textwidth]{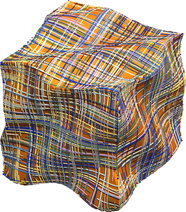}
	\end{subfigure}
	\hfill
	\\
	\begin{subfigure}[t]{0.15\textwidth}
	\includegraphics[width=1\textwidth]{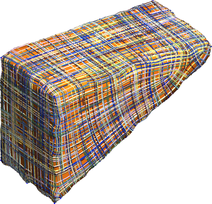}
	\end{subfigure}
	\hfill
	\begin{subfigure}[t]{0.15\textwidth}
	\includegraphics[width=1\textwidth]{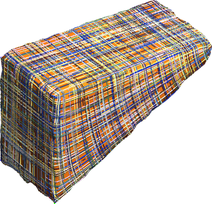}
	\end{subfigure}
	\hfill
	\begin{subfigure}[t]{0.15\textwidth}
	\includegraphics[width=1\textwidth]{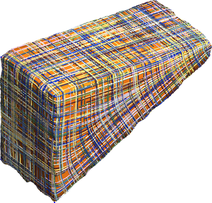}
	\end{subfigure}
	\hfill
	\begin{subfigure}[t]{0.15\textwidth}
	\includegraphics[width=1\textwidth]{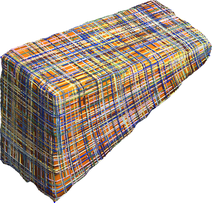}
	\end{subfigure}
	\hfill
	\begin{subfigure}[t]{0.15\textwidth}
	\includegraphics[width=1\textwidth]{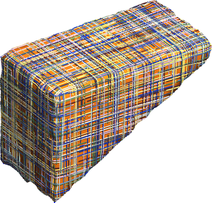}
	\end{subfigure}
	\hfill
	\begin{subfigure}[t]{0.15\textwidth}
	\includegraphics[width=1\textwidth]{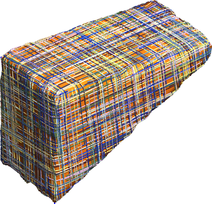}
	\end{subfigure}
	\hfill 
	\\
	\begin{subfigure}[t]{0.15\textwidth}
	\includegraphics[width=1\textwidth]{figures/fields/fandiskp/shrunk_mesh_misc_fandisk11k_n_0_p_2-streamlines.png}
	\end{subfigure}
	\hfill
	\begin{subfigure}[t]{0.15\textwidth}
	\includegraphics[width=1\textwidth]{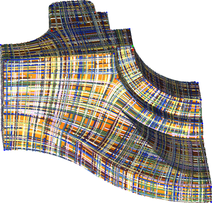}
	\end{subfigure}
	\hfill
	\begin{subfigure}[t]{0.15\textwidth}
	\includegraphics[width=1\textwidth]{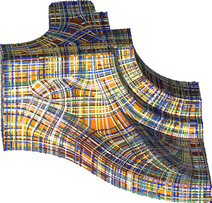}
	\end{subfigure}
	\hfill
	\begin{subfigure}[t]{0.15\textwidth}
	\includegraphics[width=1\textwidth]{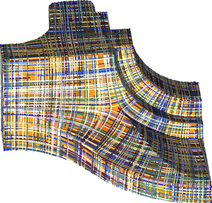}
	\end{subfigure}
	\hfill
	\begin{subfigure}[t]{0.15\textwidth}
	\includegraphics[width=1\textwidth]{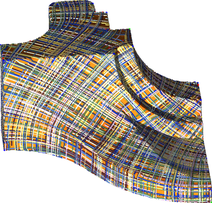}
	\end{subfigure}
	\hfill
	\begin{subfigure}[t]{0.15\textwidth}
	\includegraphics[width=1\textwidth]{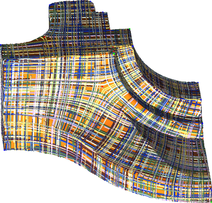}
	\end{subfigure}
	\hfill\\
	\caption{Various cross field methods compared on several meshes with complex features and geometry. We test on the three-cylinder-intersection, wavey-box, wedge, and fandisk meshes and compare against the following works: \cite{jakob_instant_2015,brandt_modeling_2018,knoppel_globally_2013} with various parameters. We use normal aligned octahedral fields generated by minimizing $E_2$. We achieve crease-alignment on all test cases where other methods succeed sporadically.}
	\label{fig:CrossFieldComparisonFig} 
\end{figure*}

\paragraph{Challenging Test Cases} We compare feature alignment of our cross fields with that of existing methods on several meshes illustrative examples in Figure~\ref{fig:CrossFieldComparisonFig}. As pointed out in the introduction, a key advantage of our technique is that it recovers crease aligned fields on models whose maximal curvature directions \emph{disagree} with their creases. This occurs naturally when models are specified by the intersections of developable patches, a very common primitive in CAD tools. We introduce two benchmark models for testing crease alignment when creases disagree with intrinsic notions of curvature. The \textbf{three-cylinder-intersection} mesh is composed of $12$ quadrilateral patches where each patch is a subset of a cylinder and has maximal curvature directions making $\frac{\pi}{4}$ angles with its boundary creases. The \textbf{wavey-box} example has the same creases as a standard cube, with the modification that each of its faces has a sine wave ripple running diagonally through it. These two cases are shown in Figure~\ref{fig:counterExamples}. The \textbf{fandisk} mesh is another example of a challenging case for feature-alignment due to its shallow crease with strong non-aligning neighboring creases which is representative of one way that such features arise in real-world models.

Our cross fields on these test cases are shown in Figure~\ref{fig:CrossFieldComparisonFig}. We observe proper feature alignment in our fields and while other methods can sometimes be tuned per model to achieve the same feature alignment, there is no choice of parameters that worked on all test cases. In particular:
\begin{itemize}
    \item 
fields from \citep{jakob_instant_2015} are distracted by extrinsic curvature on the \textbf{three-cylinder-intersection} and entirely pave over the shallow crease of the \textbf{fandisk}. Their results on \textbf{wavey-box}, and \textbf{wedge} are successfully aligned to the creases.;
\item fields from \cite{brandt_modeling_2018} are challenging to tune with $\lambda$ representing alignment to a guiding extrinsic curvature field. We show their method for the biharmonic energy ($m=2$) as a point of contrast to Dirichlet energy. We choose two values of $\lambda$, $\lambda=-.0001$ for slight extrinsic curvature alignment, and $\lambda=-.1$ for stronger extrinsic curvature alignment. Their fields are unable to align to features of the
\textbf{three-cylinder-intersection} in both cases, and specifically for $\lambda=-.1$ the field strongly aligns to noise on the flat upper face of the \textbf{fandisk} mesh. Their fields are successfully crease-aligned for the \textbf{wedge} mesh;
\item we compare against both the anti-holomorphic and Dirichlet energies of \cite{knoppel_globally_2013} with the 
curvature alignment parameter $\lambda$ set to $-0.1$. This results in good alignment on the \textbf{three-cylinder-intersection}, but noisy or unaligned fields for the remaining test cases. 
\end{itemize}
In contrast, our method for $p=2$ achieves feature alignment on all test cases without unnecessary discontinuities in the field over flat faces. We show additional results for over 200 different meshes with both smooth and creased geometries with varying values of $p$ and $\epsilon$ in the supplementary materials. The fields are crease-aligned for all creased meshes and are otherwise intrinsically smooth. For comparison, we include fields from \citep{brandt_modeling_2018} and \citep{knoppel_globally_2013} on a larger range of $\lambda$. We also include fields from \citep{brandt_modeling_2018} for $m=1$ and fields from \citep{jakob_instant_2015} in supplementary materials.

\paragraph{Quad Meshing} 

Feature alignment is especially important when using cross fields to guide high-fidelity quad meshing. 
We generate quad meshes using \citep{campen_quantized_2015} to parameterize our cross fields.
We compare against a standard quad meshing pipeline using cross fields from \citep{bommes_mixed-integer_2009} and \citep{campen_quantized_2015} for parameterization. We also test against parameterization by \citep{campen_scale-invariant_2016}, which introduces extra guidance to encourage extrinsic curvature alignment.

\begin{figure*}
    \centering
    \begin{subfigure}[t]{0.24\textwidth}
	\scalebox{-1}[-1]{\includegraphics[width=\textwidth]{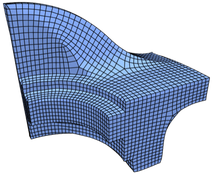}}
	\caption{$E_2$}
	\end{subfigure}%
	\hfill%
	\begin{subfigure}[t]{0.24\textwidth}
	\scalebox{-1}[-1]{\includegraphics[width=\textwidth]{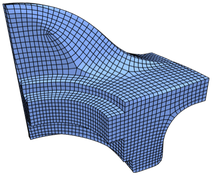}}
	\caption{$E_{\infty}$}
	\end{subfigure}%
	\hfill%
	\begin{subfigure}[t]{0.24\textwidth}
	\scalebox{-1}[-1]{\includegraphics[width=\textwidth]{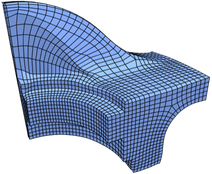}}
	\caption{QGP}
	\end{subfigure}%
	\hfill%
	\begin{subfigure}[t]{0.24\textwidth}
	\scalebox{-1}[-1]{\includegraphics[width=\textwidth]{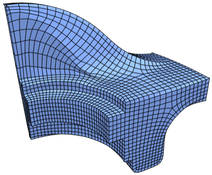}}
	\caption{Curvature filter}
	\end{subfigure}
	\caption{Quad meshes of the \textbf{fandisk} mesh generated using cross fields from $E_2$, $E_{\infty}$, MIQ + QGP \cite{campen_quantized_2015}, and MIQ + Curvature filter \cite{campen_scale-invariant_2016} respectively. Our methods achieve sharp alignment to the shallow crease with increased depth for higher $p$. Alternative methods are influenced by the crease only to a shallower extent.}
	\label{fig:QuadMeshFandiskComparison}
\end{figure*}

For the \textbf{fandisk} mesh prior methods generate quad meshes that are influenced by the shallow crease, but do not manage to capture it sharply (see Figure \ref{fig:QuadMeshFandiskComparison}). 
We observe that by placing singularities near the shallow crease of the \textbf{fandisk}, our quad meshes manage to align much more sharply. The quad mesh generated by minimizing $E_{\infty}$ aligns even better than for $E_2$.

\begin{figure*}
    \centering
    \begin{subfigure}[t]{0.24\textwidth}
	\includegraphics[width=\textwidth]{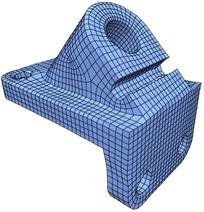}
	\caption{Anchor Mesh with $E_2$}
	\end{subfigure}%
	\hfill%
	\begin{subfigure}[t]{0.24\textwidth}
	\includegraphics[width=\textwidth]{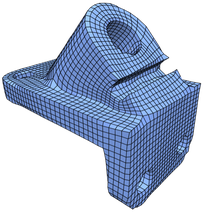}
	\caption{Anchor Mesh with \cite{bommes_mixed-integer_2009,campen_quantized_2015}}
	\end{subfigure}%
	\hfill%
	\begin{subfigure}[t]{0.24\textwidth}
	\includegraphics[width=\textwidth]{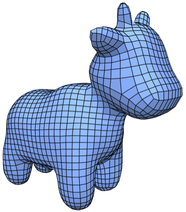}
	\caption{Spot Mesh with $E_2$}
	\end{subfigure}%
	\hfill%
	\begin{subfigure}[t]{0.24\textwidth}
	\includegraphics[width=\textwidth]{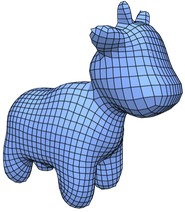}
	\caption{Spot Mesh with \cite{bommes_mixed-integer_2009,campen_quantized_2015}}
	\end{subfigure}\\
	\begin{subfigure}[t]{0.24\textwidth}
	\includegraphics[width=\textwidth]{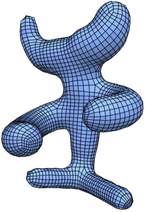}
	\caption{Moomoo Mesh with $E_2$}
	\end{subfigure}%
	\hfill%
	\begin{subfigure}[t]{0.24\textwidth}
	\includegraphics[width=\textwidth]{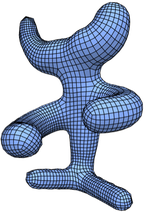}
	\caption{Moomoo Mesh with \cite{bommes_mixed-integer_2009,campen_quantized_2015}}
	\end{subfigure}%
	\hfill%
	\begin{subfigure}[t]{0.24\textwidth}
	\includegraphics[width=\textwidth]{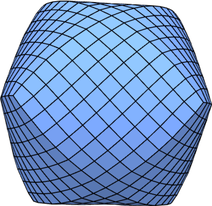}
	\caption{3 Cylinder Intersection with $E_2$}
	\end{subfigure}%
	\hfill%
	\begin{subfigure}[t]{0.24\textwidth}
	\includegraphics[width=\textwidth]{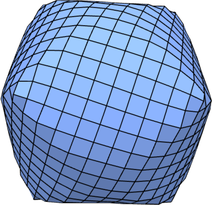}
	\caption{3 Cylinder Intersection with \cite{bommes_mixed-integer_2009,campen_quantized_2015}}
	\end{subfigure}
	\caption{Quad meshes of the \textbf{anchor}, \textbf{spot}, \textbf{moomoo}, and \textbf{three-cylinder-intersection} meshes. We compare quad meshes generated using cross fields from our $E_2$ energy with quad meshes generated through \citep{campen_quantized_2015} and \citep{bommes_mixed-integer_2009}.
	Our methods achieve sharper feature alignment on the \textbf{anchor}, \textbf{spot} (on the ear), and \textbf{three-cylinder-intersection} meshes. 
	}
	\label{fig:QuadMeshesComparison}
\end{figure*}

We also compare quad meshes generated from our cross fields against the prior art on the \textbf{anchor}, \textbf{spot}, \textbf{moomoo}, and \textbf{three-cylinder-intersection} meshes. These results are shown in Figure~\ref{fig:QuadMeshesComparison}. We observe generally better alignment in the quad meshes generated from our method. By placing singularities on the cylindrical region of the \textbf{anchor}, our quad meshing manages to align better to its creases. On the \textbf{spot} mesh we see a straighter connection between the ear and the head. For the \textbf{three-cylinder-intersection}, the quad mesh generated from our fields clearly align better. Since the \textbf{moomoo} is a relatively smooth mesh, we do not see particularly defining differences in quality.

\section{Discussion and Conclusions}

Feature alignment is a desirable property in many geometry processing applications. In the context of cross fields and remeshing, we consider features to be creases where the surface changes non-smoothly. Quality of feature detection and alignment can significantly impact quality of the remeshing and the usefulness of the resulting cross fields. While significant effort has been put into extrinsic alignment of cross fields to curvature directions, they are not always appropriate substitutes for crease alignment. By specifically targeting discontinuities of the surface we have created a new class of octahedral frame field energies parameterized by $p \geq 1$ for computing crease-aligned cross fields on surfaces. The resulting fields are intrinsically smooth over smooth surfaces, and can be used for crease-aligned quad meshing.  Moreover, alignment is fully automatic and does not rely on explicit extraction of feature curves, itself an open problem and active area of research.

We find the behavior of $E_p$ for $p\geq2$ over creases of a discretization to be an interesting point for further exploration since all practical computations on surfaces are necessarily discrete and we observe strong feature alignment, despite the problem being ill-posed in the smooth setting. Theoretical analysis of anisotropic normally-aligned octahedral frame fields combined with Proposition~\ref{prop:featureAlignment} may be able to explain this behavior. Since all edges of a mesh are creases of a piecewise linear domain, the behavior of geometry processing algorithms on creased domains merits further study.

There are also further applications of soft-normal-aligned octahedral frame fields. While in this paper we fix $\epsilon$ as a single parameter per mesh, it could also be defined as a scalar field representing ``trust'' in the quality of a mesh. It would be interesting to explore a spatially-varying $\epsilon$ dependent on triangle quality or other metrics in the future. If we treat the mesh itself as variables, soft normal alignment enables a surface flow towards meshes with lower cross-field energy. Our analysis can be further extended to SH representations of $n$-RoSy fields or even platonic solid symmetries \cite{shen_harmonic_2016}. We also conjecture that with mild assumptions the solution to our problem is unique, but a proof is out of scope for this work; we leave exploration of these ideas to future work.

Even without these extensions, our method provides a practical solution to a challenging problem. By using a new representation of cross fields we achieve crease-aligned cross fields on surfaces.

\begin{acks}
The authors would like to thank Christopher Brandt for help obtaining comparison fields, Amir Vaxman for support with Directional and rendering, Michal Adamaszek for help debugging Mosek, and David Palmer for discussions and help with his code base. We thank Wenzel Jacob, Keenan Crane, and Qingnan Zhou for their open source code implementations. Finally, we thank Ryan Viertel for many valuable discussions and Panini Pals for the panini press.

Paul Zhang acknowledges the generous support of the Department of Energy Computer Science Graduate Fellowship.
Etienne Vouga acknowledges the generous support of Adobe, SideFX, and NSF IIS-1910274.
David Bommes acknowledges the generous support of the \grantsponsor{ERC}{European Research Council (ERC)}{https://erc.europa.eu} under the European Union's Horizon 2020 research and innovation program (AlgoHex, grant agreement no.~\grantnum{ERC}{853343}). 
Justin Solomon acknowledges the generous support of \grantsponsor{ARO}{Army Research Office}{https://www.arl.army.mil/who-we-are/aro/} grant \grantnum{ARO}{W911NF-12-R-0011}, \grantsponsor{NSF}{National Science Foundation}{https://www.nsf.gov} grant \grantnum{NSF}{IIS-1838071}, \grantsponsor{AFOSR}{Air Force Office of Scientific Research}{https://www.wpafb.af.mil/afrl/afosr/} award \grantnum{AFOSR}{FA9550-19-1-0319}, and a gift from Adobe Systems. 
Any opinions, findings, and conclusions or recommendations expressed in this material are those of the authors and do not necessarily reflect the views of these organizations.

\end{acks}

\bibliographystyle{ACM-Reference-Format}
\bibliography{bibtexFile.bib}

\end{document}


\maketitle
\thispagestyle{plain}

\section{Proof of Proposition 4.1}
We carry over assumptions \S 4.1. While Equation (3) parameterizes the octahedral field by a single axis-angle rotation $v(r)$ per point $r \in \Omega$, for the case of a surface, it is more intuitive to decompose $v(r)$ into two rotations, one that accounts for intrinsic twisting, and one that accounts for the change in normal direction of the surface. 

First, we set a local parametrization of the embedding, via $\exp\!|_{r^*}: T_{r^*}\Omega \to \Omega$. WLOG, we set coordinates $(\mu,\nu)$ on $T_{r^*}\Omega$ such that axes $\mu$ and $\nu$ align with principal curvature directions (with $r^* = (0,0)$ of course).

We split $v(\mu,\nu)$ into one rotation about the normal direction $\hat{z}$ by an angle amount $t(\mu,\nu)$, and a second rotation about a vector $k(\mu,\nu) = \hat{z} \times \hat{n}(\mu,\nu)$ by angle $\arccos(\hat{z}\cdot\hat{n}(\mu,\nu))$. Note that this means $k(\mu,\nu)_z=0$. We can re-express $g(\mu,\nu)$ as the following
$$g(\mu,\nu) = e^{k(\mu,\nu) \cdot L} e^{[0, 0, t(\mu,\nu)] \cdot L} g(r^*).$$
This can be interpreted as first applying a twisting rotation about the normal, followed by a normal adjustment rotation. This implies that similar to $v(\mu,\nu)$, $k(r^*)=[0,0,0]$ and $t(r^*)=0$ as well. For shorthand, let $w(\mu,\nu) = k(\mu,\nu) + [0,0,t(\mu,\nu)]$. 

We can derive a very similar result to Equation (3) with the following changes.

\begin{equation}
\hspace{-10pt}
\label{eq:gradSPH}
\begin{aligned}
    \nabla g(\mu,\nu)|_{r^*} = \!\!
    \begin{bmatrix}
        | & | & |\\
        L_x g(r^*) & \!\!L_y g(r^*) & \!\!L_z g(r^*)\\
        | & | & |
    \end{bmatrix}
    \begin{bmatrix}
        | & | \\
        \nabla_\mu w & \!\!\nabla_\nu w \\
        | & | 
    \end{bmatrix}_{r^*}
\end{aligned}
\end{equation}

This implies that our quantity of interest the squared norm $\|\nabla g(\mu,\nu)\|^2$ at $r^*$ is
\begin{equation}
\scriptsize
\begin{aligned}
    \|\nabla g(\mu,\nu)\|^2
    =
    \Tr\left[
    \begin{bmatrix}
        | & | \\
        \nabla_\mu w & \!\!\nabla_\nu w \\
        | & | 
    \end{bmatrix}^T
    \frac{20}{3}I_3
    \begin{bmatrix}
        | & | \\
        \nabla_\mu w & \!\!\nabla_\nu w \\
        | & | 
    \end{bmatrix}
    \right]
    \\
    =
    \frac{20}{3} \Tr\left[(\nabla w)^T \nabla w\right]
\end{aligned}\label{eq:SPH_intrinsic}
\end{equation}
The $\frac{20}{3}I_3$ comes from the fact that $g L_i^T L_j g = \frac{20}{3} \delta_{ij}$ for $i,j \in {x,y,z}$ and any $g\in \mathcal{V}$\cite[Equation 3]{anonymous_algebraic_2019}
. We now expand the expression $\|\nabla w(\mu,\nu)\|_F^2$ as $|\nabla k(\mu,\nu)_x|_2^2 + |\nabla k(\mu,\nu)_y|_2^2 + |\nabla t(\mu,\nu)|_2^2$. 

Since $k(\mu,\nu)$ is an axis-angle rotation describing the change in normal from $\hat{n}(r^*)$ to $\hat{n}(\mu,\nu)$, we recognize its relationship to the principal curvatures. As $\mu$ and $\nu$ were chosen to align with principal curvature directions, $\frac{\partial k_x}{\partial \mu}=0$, $\frac{\partial k_y}{\partial \nu}=0$, $\frac{\partial k_x}{\partial \nu}=k_1$, $\frac{\partial k_y}{\partial \mu}=k_2$, where $k_1$ and $k_2$ are the principal curvatures of $\Omega$ at $r^*$. Finally denote $t_\mu$ and $t_\nu$ to be $\frac{\partial t}{\partial \mu}$ and $\frac{\partial t}{\partial \nu}$ respectively. We are left with $\|\nabla g(\mu,\nu)\|_F^2 = k_1^2+k_2^2+t_\mu^2+t_\nu^2$.

Substituting in $w$ for $t_\mu^2+t_\nu^2$, and re-expressing with Gauss and mean curvatures, K and H, we obtain $$\|\nabla g(\mu,\nu)\|_F^2 = k_1^2+k_2^2 + w = 2H^2-K+w.$$ Recall that $t(r^*)=0$. This motivates our view of $w$ as tangential intrinsic twisting.

\section{Proof of Proposition 4.2}

For more compact notation, let $\Phi \coloneqq \{ \phi \mid \forall_x |\phi(x)|_F\leq1,\phi\in C^1_c \}$. Starting from Equation (2), we may split the term into integrals over the patches $\Omega_j$:
\begin{align*}
    VTV[f] &= \sup_\Phi \sum_{i=1}^m \int_{\Omega} f_i \nabla \cdot \phi_i \\
    &= \sup_\Phi \sum_{i=1}^m \sum_{j=1}^n \int_{\Omega_j} \tilde{f}^j_i \nabla \cdot \phi_i \\
    &= \sup_\Phi \sum_{i=1}^m \sum_{j=1}^n \left(\int_{\partial \Omega_j} \tilde{f}^j_i \phi_i \cdot \hat{n} -\int_{\mathring{\Omega}_j} \nabla f_i \cdot \phi_i \right)\\
    &=  \sup_\Phi \sum_{j=1}^n \left( \left(\sum_{i=1}^m \int_{\partial \Omega_j} \tilde{f}^j_i \phi_i \cdot \hat{n} \right) - \left(\sum_{i=1}^m \int_{\mathring{\Omega}_j} \nabla f_i \cdot \phi_i \right) \right).
\end{align*}
Here, we've applied integration-by-parts and Stokes' theorem, and switched sum orders for simpler argument below. 

Let us consider the pointwise maxima over $\Omega$ for the integrands. For the integrals over $\mathring{\Omega}_j$, they are maximized for $\phi$ set to the negated, normalized $\nabla f = (\nabla f_1,\ldots,\nabla f_m)^T$. This results in an integrand value of $|\nabla f|_F$ pointwise.

The integrals over $\partial \Omega_j$ may be considered as integrals over $\Gamma$, the curve network separating the patches $\mathring{\Omega}_j$. For all but a finite set, points of $\Gamma$ lie on the interior of a curve $\gamma_k$ and separate the neighborhood into  patches. For each $\gamma_k$, choose one patch and let $f^+$ denote the limiting value from this side, and let $f^-$ denote the other limiting value. The maximizing $\phi$ will be such that $\phi_i = \left((f^+ - f^-)_i / |f^+ - f^-|_2 \right)\hat{n}^+$ where $\hat{n}^+$ is the unit normal to the $+$ patch. This results in an net integrand value of $|f^+ - f^-|_2$ once over the total curve network $\Gamma$. 

The final step is to construct a sequence of differentiable $\phi$'s that converge pointwise to these optimal values, achieving the maximum in the limit and proving our theorem. This can be done with partitions of unity, subordinate to a constructed cover. For simplicity, we assume that $\nabla f \neq 0$ on $\mathring{\Omega_j}$ and $f^+ - f^- \neq 0$ on $\gamma_k$. In this case, consider an $\epsilon > 0$ and the open cover consisting of $n$ sets $U^\epsilon_j \coloneqq \{ x | x \in \Omega_j, d(x,\partial \Omega_j) > \epsilon\}$ and $s$ sets $V^\epsilon_k \coloneqq \{ x | d(x,\gamma_k) < 2\epsilon \}$. 

On the sets $U^\epsilon_j$, we multiply the associated partition function by the negated normalized $\nabla f$. On the sets $V^\epsilon_k$, we multiply the associated partition function by any differentiable extension of $\phi|_{\gamma_k} = (\ldots,\phi_i,\ldots)$ where $\phi_i = \left((f^+ - f^-)_i / |f^+ - f^-|_2 \right)\hat{n}^+$. The sum of these is a differentiable $\phi$ that approaches the pointwise maximizers as $\epsilon \to 0$. In the case when $\nabla f$ or $f^+ - f^-$ vanishes, note that the value of $\phi$ is irrelevant at those points, and a slight modification of this argument with additional open covers (about the vanishing regions) will provide the result.

\section{Proof for Proposition 4.3}

First we derive an expression for difference between two normally aligned octahedral frames $f_0$ and $f_1$ across a crease. Let $f_0$ be the canonical frame, and $f_1=e^{b[\text{cos} a, \text{sin} a, 0]\cdot L}e^{t L_z}f_0$. The angle $t$ represents an initial twist in the $\hat{z}$ direction, and the angle $b$ represents a bend across a crease in the $xy$-plane. The crease direction is described by angle $a$ relative to $\hat{x}$. The difference between these two octahedral frames is then $E(a,b,t)=|f_0-e^{b[\text{cos}a, \text{sin}a, 0]\cdot L}e^{t L_z}f_0|^2_2.$

Consider the case where $a=0$, we can interpret $f_0$ as an octahedral frame aligned to the normal $\hat{z}$ and the crease direction $\hat{x}$. $f_1$ is then the octahedral frame on the other side of a crease of angle $\pi-b$. $t$ describes how misaligned $f_1$ is to the crease direction $\hat{x}$. The cost of deviating from crease-alignment is then $E(0,b,t)-E(0,b,0)$.

Now consider cases where $a\neq0$. Since $a$ describes the direction of the crease, its deviation from $0$ corresponds to deviation of $f_0$ from crease alignment. To prove that cost is minimized by crease alignment, it suffices to show that $E(a,b,t)-E(0,b,0)$ is non-negative for all values of $a$,$b$, and $t$.

We will first derive an expression for $E(a,b,t)$ without matrix exponentials and then use Mathematica document "FeatureAlignmentProof.nb" to prove $$E(0,b,t)-E(0,b,0) = \frac{5}{24} (7 + \cos 4 b) \;\sin^2 2t \geq 0,$$
and that $E(a,b,t)-E(0,b,0)$ is non-negative for all values of $a$, $b$, and $t$. The reason we need to first derive a form without matrix exponentials is so that Mathematica has an easier time manipulating our equations of interest. With matrix exponentials, Mathematica often hangs and is unable to output a result.

The derivation for an expression for $E(a,b,t)$ without matrix exponentials is assisted by the Mathematica document "SimpleEnergyExpression.nb" as follows. First we split an octahedral frame $f_0$ into its three lobes. $f_0 = l_x+l_y+l_z$, where $l_z=[0,0,0,0,\sqrt{\frac{7}{12}},0,0,0,0]$, $l_x=e^{\frac{\pi}{2}L_x}l_z$ and $l_y=e^{\frac{\pi}{2}L_y}l_z$. We can therefore compute $E(a,b,t)$ as follows.

\begin{equation*}
    \begin{aligned}
        E(a,b,t)&=&|f_0-e^{b[\text{cos}a, \text{sin}a, 0]\cdot L}e^{t L_z}f_0|^2_2\\
        &=&|(l_x+l_y+l_z)-e^{b[\text{cos}a, \text{sin}a, 0]\cdot L}e^{t L_z}(l_x+l_y+l_z)|^2_2\\
    \end{aligned}
\end{equation*}

Denote the rotation $e^{b[\text{cos}a, \text{sin}a, 0]\cdot L}e^{t L_z}$ by $R(a,b,t)$. This results in 

\begin{equation*}
    \begin{aligned}
        E(a,b,t)&=&|(l_x+l_y+l_z)-R(a,b,t)l_x-R(a,b,t)l_y-R(a,b,t)l_z)|^2_2\\
        &=&|l_x+l_y+l_z-R(a,b,t)l_x-R(a,b,t)l_y-R(a,b,t)l_z)|^2_2\\
    \end{aligned}
\end{equation*}

The above expression is be composed of many terms of the following two forms $l_{d_1}^T R(a,b,t) l_{d_2}$ for $d\in\{x,y,z\}$ and $|l_{d_1} - R(a,b,t) l_{d_2}|$. Note that both these expressions are variables depending on the angle between two lobes. We therefore compute $|l_{v_1}-l_{v_2}|_2^2$ and $l_{v_1}^Tl_{v_2}$ in Mathematica, where $l_v$ is a lobe oriented in the $v$ direction as a function of the angle $\theta$ between $v_1,v_2$. We obtain
$$|l_{v_1}-l_{v_2}|_2^2 = \frac{5}{42}(9+7\text{cos}2\theta)\text{sin}^2 \theta$$
and
$$l_{v_1}^Tl_{v_2} = \frac{1}{336}(9+20\text{cos}2\theta+35\text{cos}4\theta).$$
In the notebook these are denoted by "d2t[a]" and "kt[a]" respectively, where "a" in the notebook represents $\theta$. Finally we substitute these terms into $E(a,b,t)$ in "FeatureAlignmentProof.nb" to obtain $E(a,b,t)$. The final expression can be seen in the "FeatureAlignmentProof.nb". 

\section{Angular Momentum Operators}\label{sec:angularmomenumoperators}

\begin{equation*}
\begin{aligned}
L_x&=&\left(
\begin{array}{ccccccccc}
 0 & 0 & 0 & 0 & 0 & 0 & 0 & -\sqrt{2} & 0 \\
 0 & 0 & 0 & 0 & 0 & 0 & -\sqrt{\frac{7}{2}} & 0 & -\sqrt{2} \\
 0 & 0 & 0 & 0 & 0 & -\frac{3}{\sqrt{2}} & 0 & -\sqrt{\frac{7}{2}} & 0 \\
 0 & 0 & 0 & 0 & -\sqrt{10} & 0 & -\frac{3}{\sqrt{2}} & 0 & 0 \\
 0 & 0 & 0 & \sqrt{10} & 0 & 0 & 0 & 0 & 0 \\
 0 & 0 & \frac{3}{\sqrt{2}} & 0 & 0 & 0 & 0 & 0 & 0 \\
 0 & \sqrt{\frac{7}{2}} & 0 & \frac{3}{\sqrt{2}} & 0 & 0 & 0 & 0 & 0 \\
 \sqrt{2} & 0 & \sqrt{\frac{7}{2}} & 0 & 0 & 0 & 0 & 0 & 0 \\
 0 & \sqrt{2} & 0 & 0 & 0 & 0 & 0 & 0 & 0 \\
\end{array}
\right)\\
L_y&=&\left(
\begin{array}{ccccccccc}
 0 & \sqrt{2} & 0 & 0 & 0 & 0 & 0 & 0 & 0 \\
 -\sqrt{2} & 0 & \sqrt{\frac{7}{2}} & 0 & 0 & 0 & 0 & 0 & 0 \\
 0 & -\sqrt{\frac{7}{2}} & 0 & \frac{3}{\sqrt{2}} & 0 & 0 & 0 & 0 & 0 \\
 0 & 0 & -\frac{3}{\sqrt{2}} & 0 & 0 & 0 & 0 & 0 & 0 \\
 0 & 0 & 0 & 0 & 0 & -\sqrt{10} & 0 & 0 & 0 \\
 0 & 0 & 0 & 0 & \sqrt{10} & 0 & -\frac{3}{\sqrt{2}} & 0 & 0 \\
 0 & 0 & 0 & 0 & 0 & \frac{3}{\sqrt{2}} & 0 & -\sqrt{\frac{7}{2}} & 0 \\
 0 & 0 & 0 & 0 & 0 & 0 & \sqrt{\frac{7}{2}} & 0 & -\sqrt{2} \\
 0 & 0 & 0 & 0 & 0 & 0 & 0 & \sqrt{2} & 0 \\
\end{array}
\right)\\
L_z&=&\left(
\begin{array}{ccccccccc}
 0 & 0 & 0 & 0 & 0 & 0 & 0 & 0 & 4 \\
 0 & 0 & 0 & 0 & 0 & 0 & 0 & 3 & 0 \\
 0 & 0 & 0 & 0 & 0 & 0 & 2 & 0 & 0 \\
 0 & 0 & 0 & 0 & 0 & 1 & 0 & 0 & 0 \\
 0 & 0 & 0 & 0 & 0 & 0 & 0 & 0 & 0 \\
 0 & 0 & 0 & -1 & 0 & 0 & 0 & 0 & 0 \\
 0 & 0 & -2 & 0 & 0 & 0 & 0 & 0 & 0 \\
 0 & -3 & 0 & 0 & 0 & 0 & 0 & 0 & 0 \\
 -4 & 0 & 0 & 0 & 0 & 0 & 0 & 0 & 0 \\
\end{array}
\right)
\end{aligned}
\end{equation*}

\section{Detailed Runtime Data}
Detailed runtime data is provided in the "runtimeDetails.xlsx" file.

\bibliographystyle{ACM-Reference-Format}
\bibliography{bibtexFile.bib}